\documentclass[useAMS, usenatbib]{mn2e}

\usepackage{times}
\usepackage{graphicx, bm, amssymb, xcolor}
\usepackage{verbatim}
\usepackage[backref,breaklinks,colorlinks,citecolor=blue]{hyperref}
\usepackage[all]{hypcap}

\newcommand{\noopsort}[1]{}

\newcommand{\msun}{\mbox{M$_\odot$}}
\newcommand{\yr}{\mbox{${\rm yr}$}}
\newcommand{\myr}{\mbox{${\rm Myr}$}}
\newcommand{\gyr}{\mbox{${\rm Gyr}$}}
\newcommand{\pc}{\mbox{${\rm pc}$}}
\newcommand{\kpc}{\mbox{${\rm kpc}$}}
\newcommand{\mpc}{\mbox{${\rm Mpc}$}}
\newcommand{\kms}{\mbox{${\rm km}~{\rm s}^{-1}$}}
\newcommand{\cmc}{\mbox{${\rm cm}^{-3}$}}

\newcommand{\etahat}{\hat{\eta}}

\newcommand{\red}[1]{{{#1}}}

\begin{document}
\hyphenation{kruijs-sen}
\hyphenation{hera-cles}

\title[Rings and Starbursts Near Galactic Centres]{A Dynamical Model for the Formation of Gas Rings and Episodic Starbursts Near Galactic Centres}

\author[Krumholz \& Kruijssen]{Mark R. Krumholz$^1$\thanks{mkrumhol@ucsc.edu} and
J.~M.~Diederik Kruijssen$^2$\thanks{kruijssen@mpa-garching.mpg.de}
\\ \\
$^1$Department of Astronomy \& Astrophysics, University of California, Santa 
Cruz, CA 95064 USA\\
$^2$Max-Planck Institut f\"ur Astrophysik, Karl-Schwarzschild-Stra{\ss}e 1, D-85748 Garching, Germany \\
}

\date{\today}

\pagerange{\pageref{firstpage}--\pageref{lastpage}} \pubyear{2015}

\maketitle

\label{firstpage}

\begin{abstract} 
We develop a simple dynamical model for the evolution of gas in the centres of barred spiral galaxies, using the Milky Way's Central Molecular Zone (CMZ, i.e., the central few hundred pc) as a case study. We show that, in the presence of a galactic bar, gas in a disc in the central regions of a galaxy will be driven inwards by angular momentum transport induced by acoustic instabilities within the bar's inner Lindblad resonance. This transport process drives turbulence within the gas that temporarily keeps it strongly gravitationally stable and prevents the onset of rapid star formation. However, at some point the rotation curve must transition from approximately flat to approximately solid body, and the resulting reduction in shear reduces the transport rates and causes gas to build up, eventually producing a gravitationally-unstable region that is subject to rapid and violent star formation. For the observed rotation curve of the Milky Way, the accumulation happens $\sim 100$ pc from the centre of the Galaxy, in good agreement with the observed location of gas clouds and young star clusters in the CMZ. The characteristic timescale for gas accumulation and star formation is of order $10-20$ Myr. We argue that similar phenomena should be ubiquitous in other barred spiral galaxies.
\end{abstract}
\begin{keywords}
galaxies: evolution --- Galaxy: centre --- Galaxy: evolution --- ISM: kinematics and dynamics --- stars: formation
\end{keywords}

\section{Introduction}
\label{sec:intro}

Star formation is one of the most important unsolved problems in contemporary astrophysics, and one of the central uncertainties is the extent to which the rate of star formation per unit mass of interstellar gas is influenced by galactic-scale processes. (See \citealt{krumholz14c}, \citealt{dobbs14a}, and \citealt{padoan14a} for recent reviews.)  On one hand, a number of observations appear to favor a ``bottom-up" view in which star formation is a purely local process that does not depend on galactic characteristics. These include the apparent insensitivity of the star formation rates in galaxies to galactic parameters such as \citet{toomre64a} $Q$ \citep[e.g.,][]{leroy08a, leroy13a} or to the presence and structure of spiral arms \citep{willett15a}, the minimal variation in molecular gas depletion time ($t_{\rm dep}\equiv M_{\rm gas}/{\rm SFR}$) measured on $\sim \mathrm{kpc}$ scales in nearby galaxies \citep[e.g.,][]{bigiel08a, schruba11a, leroy13a}, and the fact that all star-forming systems, on scales from individual clouds to entire starburst galaxies, appear to turn their mass into stars at a nearly constant rate per gas free-fall time \citep{krumholz07e, krumholz12a}\red{, possibly with a small secondary variation based on the Mach number of the turbulence \citep{federrath13c}}. In this view, the global rate of star formation in galaxies is mostly a matter of adding up individual star-forming clouds, whose behavior is independent of their galactic environment. Models based on this premise have been advanced by a number of authors \citep[e.g.,][]{krumholz05c, krumholz09b, lada12a, renaud12a, krumholz13c, federrath13c, salim15a}.

On the other hand, there is a broad class of theoretical models predicts that galaxies' star formation rates should explicitly depend on galactic structure, at least in some galaxies \citep[e.g.,][]{thompson05a, ostriker10a, ostriker11a, shetty12a, hopkins11a, faucher-giguere13a}, and a number of recent observations using higher resolution data have pointed to a somewhat greater role for galactic dynamics. These include observations showing that the properties of molecular clouds vary systematically with environment both between and within galaxies \citep[e.g.,][]{hughes13a, colombo14a}, and evidence that the star formation rate per unit molecular mass measured on small scales is not independent of the large-scale rate of shear in a galactic disc \citep{meidt13a, suwannajak14a}.

A particularly promising avenue for exploring the role of galaxy-scale dynamics in regulating star formation is to study the centres of galaxies. These are the regions where shear and similar effects arising from the shape of the galactic potential should have their largest effects \citep[e.g.,][]{kruijssen14b, kruijssen15a}. It is also the location where other environmental effects, such as high external pressures \citep[e.g.,][]{rathborne14a} and high X-ray and cosmic ray fluxes \citep[e.g.,][]{meijerink05a, papadopoulos10a, meijerink11a, clark13b, kruijssen14b}, should be at their strongest. Indeed, galactic centres show a number of interesting deviations from the star formation behavior seen at larger galactic radii. Some galaxies show enhanced star formation per unit molecular mass in their nuclei, others depressed rates of star formation \citep{saintonge12a, leroy13a,longmore13a}. Galactic centre molecular clouds show systematic differences in their properties from disc clouds \citep[e.g.,][]{kruijssen13a,leroy14a, rathborne14a,rathborne15a,bally14a}.

The best-studied of galactic centres is the Central Molecular Zone (CMZ) of the Milky Way, which exhibits a number of interesting features. Much of the dense gas in this region appears to be collected into a stream, or a partially filled ring, of molecular material $\sim 100$ pc from the Galactic centre \citep{molinari11a, kruijssen15a}. Within this ring are a series of clouds whose star forming activity ranges from far smaller than one would expect based on their mass and density \citep[e.g., G0.253+0.016, also known as ``The Brick";][]{longmore12a,kauffmann13a,rathborne14b,mills15a} to some of the most actively star-forming sites in the Local Group \citep[Sgr A, Sgr B2, and Sgr C;][]{yusef-zadeh08a, yusef-zadeh09a}. These clouds may well represent an evolutionary sequence \citep{longmore13b,kruijssen15a}. There is also significant evidence that star formation in the CMZ is episodic. Hints at previous starbursts include the presence of large off-plane bubbles visible in radio \citep{sofue84a}, infrared \citep{bland-hawthorn03a}, and gamma-rays \citep{su10a}, along with direct counts of young stellar objects \citep{yusef-zadeh09a}.

In this paper we propose a simple model for the global dynamics and star formation behaviour of gas in the Milky Way's CMZ, and by extension the central regions of other barred spiral galaxies. Qualitatively, our model is that gas is channelled through the disc of the Milky Way, along the Galactic bar, and into the CMZ. Once there, it settles into a disc that is subject to acoustic instabilities driven by the bar. These instabilities drive both turbulence and angular momentum transport in the gas, simultaneously causing it to flow inward and increase in velocity dispersion. As a result, the gas is extremely turbulent, and its large velocity dispersion renders it highly stable against gravitational collapse or star formation (typically $Q\sim 100$, as we show below). However, near the radius where the rotation curve of the Galaxy turns over from flat to solid body, the resulting reduction in shear suppresses transport and turbulent driving, leading gas to accumulate rather than moving further inward. This accumulation eventually builds up a ring of material which goes gravitationally unstable and begins vigorous star formation. If feedback from this burst of star formation is sufficiently strong, the ring will be disrupted, and the cycle will begin again.

Our plan for the remainder of this paper is as follows. In \autoref{sec:model} we introduce our physical model for the gas in the CMZ, and in \autoref{sec:simulations} we describe how we simulate the evolution of this model. In \autoref{sec:results} we present the results of our simulations, and we discuss the implications of those results in \autoref{sec:discussion}. We summarize and conclude in \autoref{sec:conclusion}.

\section{Model}
\label{sec:model}

Our goal in this section is to develop a quantitative model corresponding to the qualitative scenario described in the previous section. We will model the gas near a galactic centre as an axisymmetric thin disc subject to non-axisymmetric perturbations, within which mass and angular momentum are transported via instabilities and energy is lost due to the dissipation of supersonic turbulence. We will simulate this system using the \texttt{VADER} code of \citet{krumholz15a}. In the remainder of this section we detail the physical ingredients to our model, and in the following one we describe the numerical setup we use to simulate it. Our simulations will focus on the case of the Milky Way CMZ, since it is the system for which we have the best measurements of the small-scale rotation curve.

\subsection{The Galactic Potential and the Inner Lindblad Resonance}
\label{ssec:potential}

\begin{figure}
\includegraphics[width=\columnwidth]{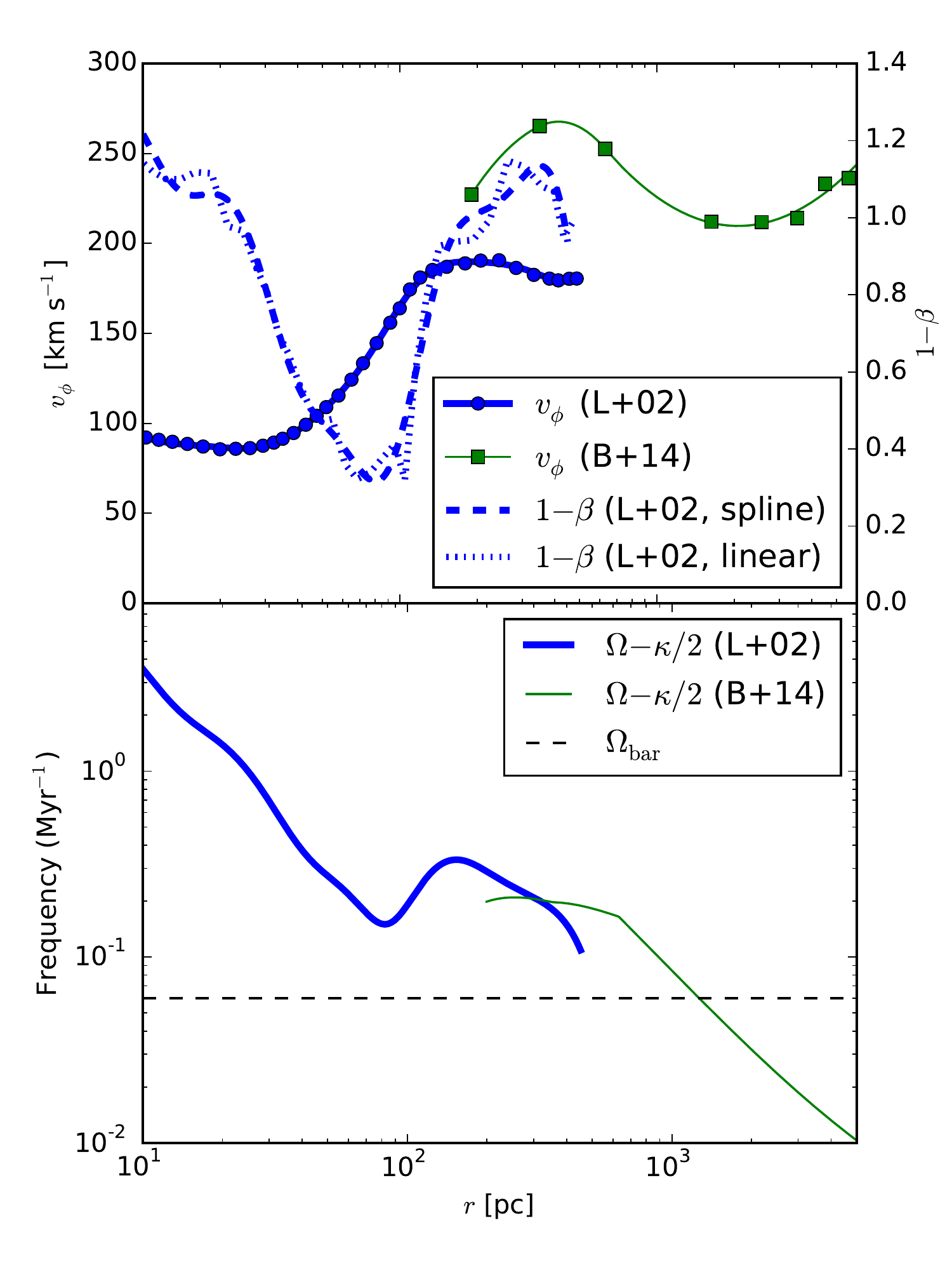}
\caption{
\label{fig:rotcurve}
Rotation curve for the inner Galaxy. \textit{Top panel:} rotation speed $v_\phi$ versus Galactocentric radius $r$ using the data set of \citet[\textit{blue circles and thick solid line}]{launhardt02a} and of \citet[\textit{green squares and thin solid line}]{bhattacharjee14a}. For the $v_\phi$ versus $r$ plots, the points indicate the observational estimate, while the lines are B-spline fits to the data (see text for details). The \textit{dashed blue line} shows the dimensionless shear $1-\beta=1-d\ln v_\phi/d\ln r$ derived from the B-spline fit to the \citeauthor{launhardt02a} data; for comparison, the \textit{dotted blue line} shows $1-\beta$ computed using a simple linear fit. \textit{Bottom panel:} angular frequencies versus radius. \textit{Solid lines} show $\Omega_0-\kappa/2$, where $\Omega_0$ is the orbital frequency and $\kappa$ is the epicyclic frequency, for the fits to the \citeauthor{launhardt02a} (\textit{thick blue line}) and \citeauthor{bhattacharjee14a} (\textit{thin green line}) data sets. The \textit{thin black dashed line} shows the pattern speed of the Galactic bar; the intersection of this line with $\Omega_0-\kappa/2$ is location of the ILR.
}
\end{figure}

The first step in this modeling is to derive the potential in which the gas orbits. To do so, we make use of two data sets for the Milky Way's CMZ. Our primary data set, which we use for all numerical calculations, is the enclosed mass $M_r$ versus galactocentric radius $r$ measured by \citet{launhardt02a} at $r=0.63-488$ pc, which we use to derive a rotation speed $v_\phi=\sqrt{GM_r/r}$. Our secondary data set, which we will not use for numerical computations but that we include here because the \citet{launhardt02a} data do not go out far enough to reach the inner Lindblad resonance \citep[ILR; e.g.,][]{binney87a}, is the rotation curve compiled by \citet{bhattacharjee14a} from $r=190-1.9\times 10^5$ pc.\footnote{\citet{bhattacharjee14a} provide three possible fits, corresponding to three choices for the imperfectly known Galactocentric radius of the Solar Circle. We use their data set corresponding to 8.0 kpc for this value, but the results are qualitatively identical for other choices.} We can interpolate these tabulated data to generate a rotation curve $v_\phi$ versus $r$, but this requires some care. The dynamical evolution of gas in this potential depends not just on the rotation curve but on its gradient,
\begin{equation}
\beta \equiv \frac{d\ln v_\phi}{d\ln r}.
\end{equation}
This quantity is important because the dimensionless rate of shear is $1-\beta$. For this reason we interpolate using basis splines (B-splines), following the method of \citet{gans84a}, as implemented in the \texttt{VADER} code by \citet{krumholz15a}. The B-spline fit ensures continuity of some number of derivatives. For the \citet{launhardt02a} data set we use 6th order B-splines with 15 breakpoints, while for the \citet{bhattacharjee14a} data set, since it contains fewer data points within 6 kpc (where we truncate the fit), we use 3rd order B-splines with 6 breakpoints. The top panel of \autoref{fig:rotcurve} shows the data and our fits to them. We see that there is some tension between the two data sets, but that there is rough qualitative agreement in the range of radii where they overlap. The Figure also illustrates why the B-spline fit is critical: a simple linear fit produces rates of shear with unphysical sharp features.

The lower panel of \autoref{fig:rotcurve} shows, for our fits to both data sets, the frequency $\Omega - \kappa/2$, where $\Omega=v_\phi/r$ is the angular velocity of the orbit and $\kappa = \sqrt{2(1+\beta)}\Omega$ is the epicyclic frequency. For comparison, the black dashed line shows the pattern speed $\Omega_{\mathrm{bar}} \approx 0.06$ Myr$^{-1}$ of the Galactic bar \citep{debattista02a, wang12b, antoja14a}; the point where $\Omega - \kappa/2 = \Omega_{\mathrm{bar}}$ marks the location of the ILR. We see that the ILR is located at $\sim 1$ kpc. The entire region we model will be within this radius.

Finally, note that, on top of this cylindrically-symmetric potential, there is an additional non-axisymmetric component to the potential due the presence of a bar. We incorporate this effect into our model below by considering how the presence of a non-axisymmetric perturbation affects this disc, and so we defer any further considering of the bar's effects for the moment.

\subsection{Gas Evolution}

We approximate that gas in the CMZ orbits in an azimuthally-symmetric thin disc, which is characterized by a surface density $\Sigma$, a non-thermal velocity dispersion $\sigma_{\mathrm{nt}}$, and a thermal velocity dispersion $\sigma_{\mathrm{th}}$. Both $\Sigma$ and $\sigma_{\mathrm{nt}}$ are functions of $r$, but we set $\sigma_{\mathrm{th}}=0.5$ km s$^{-1}$ independent of position. This is the sound speed of fully molecular gas at a temperature of 70 K, in the middle of the kinetic temperature range for CMZ gas that \citet[also see \citealt{ginsburg15a}]{ao13a} derive by analyzing H$_2$CO line emission. The exact choice will have little effect on our results in any event, because observations show, and the model we describe below also predicts, that $\sigma_{\mathrm{nt}} \gg \sigma_{\mathrm{th}}$ almost everywhere in the disc.

The gas evolves following the standard equations of mass and energy conservation for such a disc \citep[e.g.,][]{krumholz10c},
\begin{eqnarray}
\label{eq:masscons}
\frac{\partial}{\partial t}\Sigma + \frac{1}{r} \frac{\partial}{\partial r} (r v_r \Sigma) & =& 0 \\
\label{eq:encons}
\frac{\partial}{\partial t} E + \frac{1}{r}\frac{\partial}{\partial r}\left[rv_r\left(E+P\right)\right] -
\frac{1}{r}\frac{\partial}{\partial r}\left(r \frac{ v_\phi \mathcal{T}}{2\pi r^2}\right) & = & \dot{E}_{\mathrm{rad}},
\end{eqnarray}
where $E = \Sigma\left[\psi + v_\phi^2/2 + (3/2)\left(\sigma_{\mathrm{nt}}^2+\sigma_{\mathrm{th}}^2\right)\right]$ is the total energy of the gas per unit area, $\psi$ is the gravitational potential, $\dot{E}_{\mathrm{rad}}$ is the rate of energy gain or loss due to radiative effects, $P = \Sigma\left(\sigma_{\mathrm{nt}}^2+\sigma_{\mathrm{th}}^2\right)$ is the total vertically-integrated pressure, $v_r$ is the radial velocity, and $\mathcal{T}$ is the turbulent torque. The latter two quantities are related via angular momentum conservation,
\begin{equation}
v_r = \frac{\partial \mathcal{T}/\partial r}{2\pi r\Sigma v_\phi (1 + \beta)}.
\end{equation}
We parameterize the viscosity though the \citet{shakura73a} $\alpha$-disc model, including the generalization proposed by \citet{shu92a}:
\begin{equation}
\label{eq:torque}
\mathcal{T} = -2\pi r^2 \alpha P \left(1 - \beta\right),
\end{equation}
where $\alpha$ is a dimensionless number characterizing the strength of angular momentum transport. We defer a discussion of the choice of $\alpha$ to the next section.

It is worth pausing to remark on the physical processes that are responsible for controlling the turbulence in the gas that are captured in \autoref{eq:encons}, since these will be crucial to understanding our results. The non-thermal velocity dispersion $\sigma_{\mathrm{nt}}$ is a component of the total energy per unit mass $E/\Sigma$. Since all the other components depend on quantities that are fixed at each radius ($\psi$, $v_\phi$, and $\sigma_{\mathrm{th}}$), any change in the energy $E$ at a given location is expressed as a change $\sigma_{\mathrm{nt}}$. These changes can be driven by several processes, each corresponding to a different term in \autoref{eq:encons}. The easiest to understand is the decay of supersonic turbulence via radiative shocks, which is captured by the term $\dot{E}_{\mathrm{rad}}$. This process causes $\sigma_{\mathrm{nt}}$ to decrease at every point in the disc.

The countervailing processes are represented by the terms on the left-hand side of \autoref{eq:encons}. The term $-(1/r)(\partial/\partial r)(r v_\phi \mathcal{T}/2\pi r^2)$ describes the work done by torques. Physically, in our viscous disc model angular momentum transport occurs because each ring of material has a higher angular velocity than the ring immediately outside it, and as the fluid elements move past one another they exert torques. This torque does work, the effect of which is to transport energy outward. Finally, the term $(1/r)(\partial/\partial r)[r v_r(E+P)]$ combines advection of energy by bulk motion of gas and work done by pressure forces. Since $E\gg P$ for a thin disc, advection is by far the more important of these two processes. Advection tends to increase $\sigma_{\mathrm{nt}}$ because, although fluid elements retain constant specific energy $E/\Sigma$ as they advect, the gravitational potential $\psi$ and rotation curve $v_\phi$ are such that $\psi+v_\phi^2/2$ decrease inward. Thus the specific orbital plus gravitational energy of a fluid element goes down as it is transported inward, and by energy conservation this leads the specific turbulent energy $(3/2)\sigma_{\mathrm{nt}}^2$ to increase. Thus the velocity dispersion in our disc is dictated by a competition between the decay of turbulence in radiative shocks, which decreases $\sigma_{\mathrm{nt}}$, and the combined effects of turbulent torques and advection, which tend to increase it by trading off gravitational potential energy against kinetic energy.

\subsection{Energy Dissipation Processes}

The next step in our model is to evaluate $\dot{E}_{\mathrm{rad}}$. In the absence of any forcing, the non-thermal velocity dispersion of the gas will decay with time. Numerical simulations indicate that the timescale for turbulence to decay is of order the crossing time of the flow \citep{stone98a, mac-low99b, ostriker01a, lemaster09a}. In the context of a disc, this timescale should be computed relative to the gas scale height, which we obtain using the model of \citet{ostriker10a} for the vertical gas structure. In this model, the scale height $H_g$ of the gas is implicitly given by
\begin{equation}
\label{eq:scaleheight}
2\pi \zeta_d G \rho_* \Sigma H_g^2 + \frac{\pi}{2}G \Sigma^2 H_g = P,
\end{equation}
where $\rho_*$ is the stellar (plus dark matter) mass density within a distance $H_g$ of the gas midplane (assumed to vary negligibly, which is true as long as the stellar scale height greatly exceeds $H_g$) and $\zeta_d\approx 0.33$ is a dimensionless constant whose exact value depends on the relative importance of gas self-gravity versus stellar gravity, but whose value remains within 5\% of $0.33$ in all cases. Intuitively, \autoref{eq:scaleheight} simply asserts that, in hydrostatic balance, the vertically-integrated pressure must balance the gravitational forces attempting to compress the gas, which come from the gravitational potential produced by stars and dark matter (the first term on the left hand side) and the self-gravity of the gas (the second term).

To obtain the stellar density, we note that, if the stellar mass responsible for producing the rotation velocity $v_\phi$ were arranged spherically, we would have
\begin{equation}
\rho_* = f_{\mathrm{shape}} (1+2\beta) \frac{v_\phi^2}{4\pi G r^2}
\end{equation}
with $f_{\mathrm{shape}} = 1$. A flattened stellar distribution has $f_{\mathrm{shape}}>1$. We do not precisely know the full three-dimensional distribution of mass at the Galactic Centre, but $f_{\mathrm{shape}}$ can be constrained somewhat by observations. \citet{rodriguez-fernandez08a}, based on 2MASS star counts, and \citet{molinari11a}, based on the shape of the 100 pc gas structure in the CMZ, both estimate $f_{\mathrm{shape}}\sim 2$. More recently, \citet{kruijssen15a} fit the potential based on the kinematics of gas in the CMZ\footnote{\citet{kruijssen15a} express their result in terms of the vertical shape of the potential, expressed via the axial ratio of the equipotential surfaces $q_\Phi$ \citep[section 2.2.2]{binney87a}. Their favored value is equivalent to $f_{\mathrm{shape}} = 2.5$.}, and obtained $f_{\mathrm{shape}}\approx 2.5$. We will adopt this values as our fiducial choice, but also study how varying it affects our results.

Given a choice for $f_{\mathrm{shape}}$, at any point in the disc with known $\Sigma$, $P$, and $v_\phi$, \autoref{eq:scaleheight} is a quadratic with known coefficients that we can solve to obtain $H_g$. We then set the energy dissipation rate to
\begin{equation}
\label{eq:cooling}
\dot{E}_{\mathrm{rad}} = -\eta \frac{\Sigma \sigma_{\mathrm{nt}}^2}{H_g/\sigma_{\mathrm{nt}}},
\end{equation}
where $\eta$ is a constant of order unity, for which we take a fiducial value of 1.5, calibrated from simulations \citep{stone98a}. This amounts to setting the rate of energy loss equal to the current kinetic energy divided by one gas scale-height crossing time.

\subsection{Angular Momentum Transport Processes}

The final important element in our model is angular momentum transport, as parameterized by the dimensionless value $\alpha$. To estimate this quantity, consider a disc of molecular gas in a barred galaxy, such as that in the CMZ. The stellar bar will perturb the gas disc with an $m=2$ mode, and we wish to consider the stability of the disc against these perturbations. Stability analyses such as this have been performed by a large number of authors \citep[e.g.,][]{goldreich65a, lau78a, toomre81a, bertin89a, montenegro99a}. \citet{bertin89a} obtained the general dispersion relation for non-axisymmetric self-gravitating thin discs, and \citet{montenegro99a} pointed out that it admits two classes of instability: gravitational instabilities, which occur for axisymmetric perturbations and for non-axisymmetric ones close to co-rotation, and acoustic instabilities, which arise strictly for non-axisymmetric perturbations inside the ILR or outside the outer Lindblad resonance (OLR). Acoustic instabilities are so named because the destabilization is driven by pressure rather than gravity; pressure causes the apocentres of perturbed gas orbits inside the ILR to align, leading to a growing mode. Indeed, gravity acts as a stabilizer rather than a destabilizer for acoustic instabilities, since it ``de-tunes" the pressure-driven alignment.

Formally, the dispersion relation for waves in a self-gravitating thin disc is \citep[their Equation 10]{montenegro99a}
\begin{equation}
\frac{Q^2}{4} = \etahat - \frac{1-\nu^2}{\etahat^{-2}+J^2/(1-\nu^2)}.
\label{eq:dispersion}
\end{equation}
where $\nu = (\omega - m\Omega)/\kappa$ is the dimensionless frequency of the perturbation, $\omega$ is the frequency, $m$ is the azimuthal wavenumber of the perturbation, and $\etahat$ is a dimensionless inverse wavenumber, which must be positive and real. The instability is controlled by the two parameters $Q$ and $J$, where
\begin{equation}
Q = \frac{\kappa \sigma}{\pi G \Sigma}
\end{equation}
is the usual \citet{toomre64a} parameter, and
\begin{equation}
J = \frac{\sqrt{T_1}}{k_{\mathrm{crit}}},
\end{equation}
with $T_1 = (1-\beta)(2 m \Omega/\kappa r)^2$ and $k_{\mathrm{crit}} = \kappa^2 / 2\pi G\Sigma$, is a dimensionless measure of the strength of shear in the disc. Note that $J = 0$ for $\beta=1$, corresponding to solid body rotation and thus zero shear. The disc is unstable to the growth of perturbations with azimuthal wavenumber $m$ at a given combination of $Q$ and $J$ if there exists a positive real value of $\etahat$ such $\nu$ has a non-zero imaginary part.

Equation (\ref{eq:dispersion}) can be solved for $\nu^2$ as
\begin{equation}
\nu^2 =  1 + \frac{1}{2}\left(\frac{Q^2}{4 \etahat^2} - \frac{1}{\etahat} \pm \sqrt{D}\right),
\label{eq:dispersion1}
\end{equation}
with
\begin{equation}
D = \left(\frac{Q^2}{4\etahat^2}-\frac{1}{\etahat}\right)\left(\frac{Q^2}{4\etahat^2}-\frac{1}{\etahat}-4 J^2\etahat^2\right).
\end{equation}
However, be warned that this re-arrangement introduces some spurious solutions, as is often the case with rational equations. Nonetheless, this re-arrangement is useful, because it clearly shows how to analyze a disc to determine whether it is either gravitationally or acoustically unstable.

The two regimes of instability correspond to different signs of the discriminant $D$. The gravitational instability case arises when $D>0$, so $\nu^2$ is a real number, but $\nu^2 < 0$ so that $\nu$ is a pure imaginary.  In this case the transition from stability to instability occurs in the vicinity of $\nu = 0$, which corresponds to unstable waves being close to co-rotation with the disc ($\omega \approx m\Omega$). For $m=J=0$, corresponding to axisymmetric perturbations, the dispersion relation (\autoref{eq:dispersion}) reduces to
\begin{equation}
\nu^2 = 1 + \frac{Q^2}{4\etahat^2} - \frac{1}{\etahat},
\end{equation}
which has a minimum of $\nu^2 = 1-Q^{-2}$ at $\etahat=Q^2/2$. This minimum is negative, indicating instability, if $Q < 1$, the usual \citet{toomre64a} stability condition. For non-axisymmetric perturbations, $m\neq 0$, but in real astrophysical systems we generally have $J \ll 1$ (see \citealt{bertin89a} for more discussion of this point), so this condition is modified only slightly. Acoustic instability corresponds to the regime where $D$ is itself negative, and thus $\nu^2$ and $\nu$ both have non-zero imaginary parts. In this case the real part of $\nu$ is either $<-1$ or $>1$, and thus in the portion of the disc that lies either inside the ILR or outside the OLR with respect to the perturbation.

For practical numerical purposes, we can evaluate the susceptibility of a disc with specified values of $\Sigma$, $\sigma$, and rotation curve $v_\phi$ to acoustic instability via the following procedure. First, we find the value of $\etahat$ that minimizes $D$; the global minimum of $D$, if it is less than zero, corresponds to the fastest growing acoustic mode. Local minima of $D$ occur at values of $\etahat$ that satisfy $dD/d\etahat=0$, and evaluating the derivative gives
\begin{equation}
16 J^2 \etahat^5 - 8 \etahat^2 + 6 Q^2 \etahat - Q^4 = 0.
\end{equation}
We obtain the roots of this polynomial via standard numerical techniques \citep{galassi09a}, discard roots where $\mathrm{Re}(\etahat) < 0$ or $\mathrm{Im}(\etahat)\neq 0$ (since only positive real roots are physically allowed), and check the remaining roots to see which one minimizes $D$. If $D$ is negative at its minimum, the region is acoustically unstable, and we can obtain the growth rate of the most unstable mode simply by plugging this minimum into \autoref{eq:dispersion} and finding which of the four possible values of $\nu$ has the largest imaginary part.

\begin{figure}
\includegraphics[width=\columnwidth]{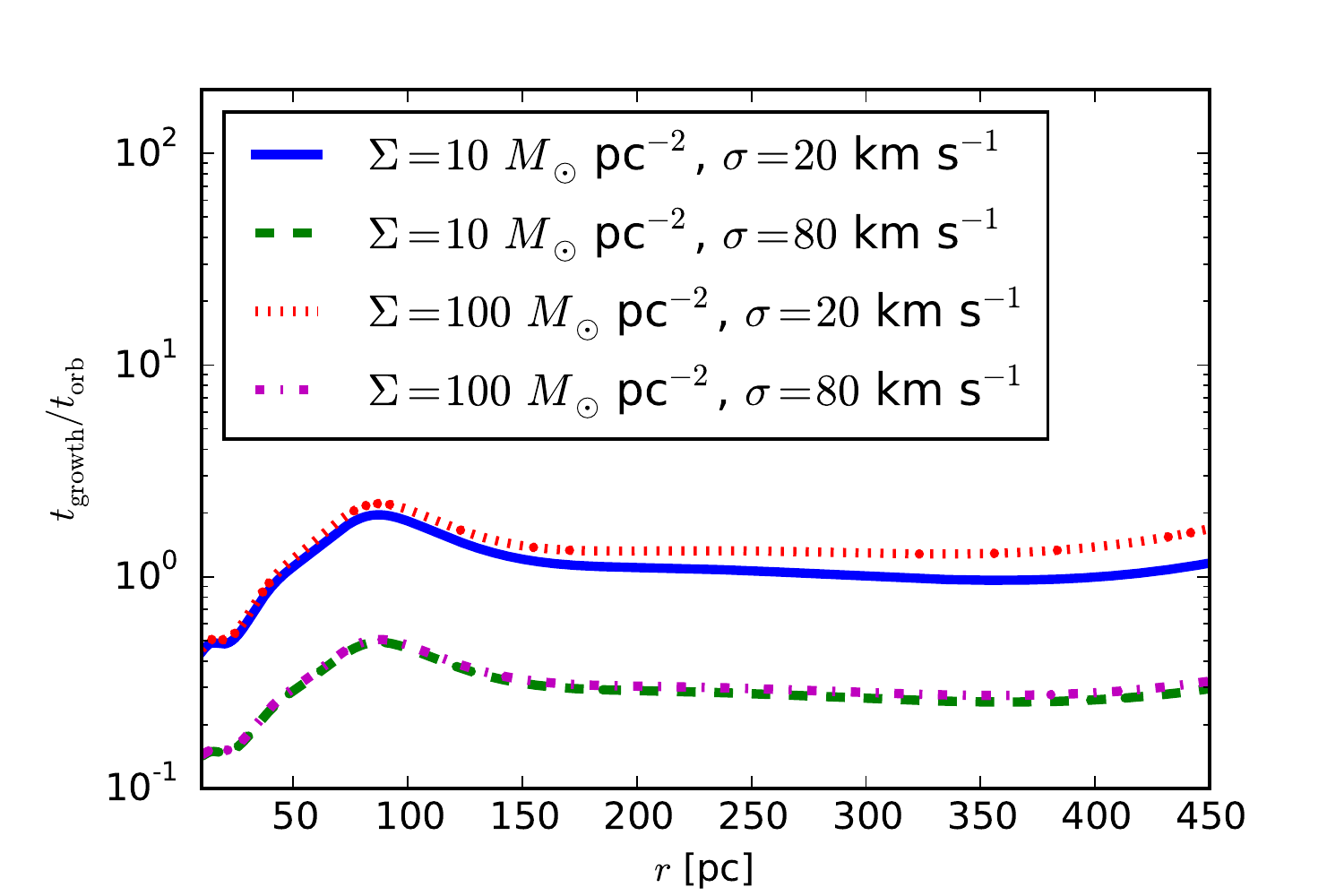}
\caption{
\label{fig:tgrowth}
Growth timescale $t_{\mathrm{growth}}$ for the fastest-growing acoustic instability mode at the Galactic Centre, normalized to the local orbital period $t_{\mathrm{orb}} = 2\pi/\Omega$, as a function of Galactocentric radius $r$. The growth timescales shown have been computed for values of the surface density $\Sigma=10$ and 100 $M_\odot$ pc$^{-2}$ and velocity dispersion $\sigma=20$ and $80$ km s$^{-1}$, independent of position, and for $m=2$ modes. The rotation curve used in the computation is produced by B-spline interpolation of the \citet{launhardt02a} measurements following the procedure outlined in \autoref{ssec:potential}, as shown by the thick blue solid line in the upper panel of \autoref{fig:rotcurve}.
}
\end{figure}

Formally, we define the growth time of the fastest growing mode as
\begin{equation}
t_{\mathrm{growth}} = \frac{1}{\kappa \max[\,\mathrm{Im}(\nu)]},
\end{equation}
where in evaluating possible values of $\nu$ we consider both gravitational and acoustic instabilities. To investigate the stability of the Milky Way's CMZ, we use the \citet{launhardt02a} potential, again treating the potential as cylindrically-symmetric, and regarding the much smaller non-axisymmetric component provided by the bar as a perturbation. For this potential, we compute the stability of the disc against both $m=0$ gravitational instabilities and $m=2$ acoustic perturbations driven by the Galactic bar, using a range of sample values of $\Sigma$ and $\sigma$. We show the results of this computation in \autoref{fig:tgrowth}. We find that, for the parameter choices shown, which span the plausible range based on observations, the disc is unstable to acoustic modes at all radii; it is not unstable to gravitational modes anywhere, although, as we will see below, it will naturally evolve into a gravitationally unstable state. The growth times of the acoustic modes are generally comparable to the orbital period, indicating that the instability should grow efficiently. The exception is just inside $\sim 100$ pc, where the rotation curve approaches solid body and the shear diminishes (c.f.~\autoref{fig:rotcurve}). This will prove to be important below.

The presence of an instability will certainly produce turbulence and angular momentum transport. Unfortunately, we cannot easily compute the exact transport rate once the instability reaches full non-linear saturation. However, it seems likely that they will be high. Simulations of both pure gas discs and gas plus stellar discs show that gravitational instabilities tend to produce transport rates corresponding to a dimensionless viscosity $\alpha\sim 1$ \citep[e.g.,][]{bournaud07a, kratter10a, hopkins10b, ceverino15a}. Acoustic instabilities, unlike gravitational ones, cannot self-stabilize by producing turbulence and thus driving up the value of $Q$ \citep[e.g.,][]{krumholz10c}, so the instability seems likely to be at least as strong. It is likely that the growth of the instability is ultimately limited by the dissipation of energy in spiral shocks.

Given the lack of a well-motivated estimate for the transport rate due to acoustic instability, we choose to parameterize it by
\begin{equation}
\label{eq:alphaparam}
\alpha = \min(\alpha_0 e^{1-t_{\mathrm{growth}}/t_\mathrm{orb}}, 1)
\end{equation}
where $t_{\mathrm{orb}} = 2\pi/\Omega$ is the local orbital period. The parameter $\alpha_0$ simply normalizes the rate of angular momentum transport at points where the instability grows on a timescale of a single orbital period, while taking the minimum ensures that, even in the most unstable regions, we do not exceed $\alpha = 1$. We will adopt a fiducial value $\alpha_0 = 1$, but consider below how varying this parameter might change the results.

\subsection{Summary of the Model}

We now have in place a fully-specified model of the evolution of gas in the Milky Way's CMZ. The overall evolution is described by \autoref{eq:masscons} and \autoref{eq:encons}, which specify mass and energy conservation. These equations depend on the rotation curve $v_\phi$ and potential $\phi$, the rate of energy dissipation $\dot{E}_{\mathrm{rad}}$, and the dimensionless rate of angular momentum transport $\alpha$. For the rotation curve and potential we use the B-spline fit to the \citet{launhardt02a} rotation curve shown in \autoref{fig:rotcurve}. For the rate of radiative energy loss we use the value given by \autoref{eq:cooling}, which comes from our model of turbulent dissipation. Finally, for angular momentum transport we use our calculation of acoustic and gravitational instabilities, as parameterized by \autoref{eq:alphaparam}. This set of equations, once we choose values for the various parameters appearing in them, fully specifies the problem.

\section{Simulation Method and Setup}
\label{sec:simulations}

We simulate the disc model described in the previous section using the \texttt{VADER} code described by \citet{krumholz15a}.\footnote{We make one minor modification to the system as described: we adopt a minimum viscosity $\alpha=10^{-3}$. We do this both because very small values of $\alpha$ cause difficulties in convergence that slow the calculations, and because we expect $\alpha$ to be at least this large as a result of thermal and magneto-rotational instability \citep[e.g.][]{piontek04a, piontek07a}, even in regions that are acoustically and gravitationally stable. The value of the floor parameter does not affect the results.} \texttt{VADER} solves the equations using a fully-implicit method that conserves mass and energy to machine precision. We use \texttt{VADER}'s backwards-Euler solver, because the Crank-Nicolson one produces undesirable numerical oscillations. All our simulations use a computational grid of 512 cells, spaced logarithmically from an inner edge at $r=10$ pc to an outer edge at $r=450$ pc. Although the region inside the ILR extends to $\sim 1$ kpc, and the bar goes out even further, we choose 450 pc for the outer edge because it is close to the outermost point included in \citet{launhardt02a}'s estimate of the Galactic rotation curve, and thus avoids the need to interpolate between this rotation curve and that of \citet{bhattacharjee14a}. We have verified that choosing different values that are still of order hundreds of pc does not qualitatively alter the results.

The equations describing the evolution of the disc require two boundary conditions at each end, describing the rate at which mass and enthalpy are advected into or out of the computational domain. At the inner boundary, we prescribe our boundary by requiring that the viscous torque go to zero. This implicitly sets the mass flux (see \citealt{krumholz15a} for details), and forces it to be out of the computational domain. We set the specific enthalpy at the inner boundary equal to the initial specific enthalpy in the simulation (see below), but this has no practical effect since no material ever enters the computational domain at the inner edge.

The choice of outer boundary condition is less trivial. We choose to specify the boundary condition as a fixed inward mass flux, but we do not have good observational estimates of the rate at which material is transported through the outer parts of the Milky Way's disc to enter the CMZ, or of the velocity dispersion of this material as it arrives. Theoretical estimates suggest that the inward transport rate through the disc should be of order $\dot{M}_{\mathrm{in}} = 1$ $M_\odot$ yr$^{-1}$ \citep{krumholz10c, forbes12a, forbes14a, cacciato12a}, and that, once gas reaches the bar, it should be further transported inward along the bar to be deposited in the CMZ \citep{binney91a, kormendy04a}\red{, where it settles into the disk-like structure that we seek to model \citep{sormani15a}.} We therefore adopt $\dot{M}_{\mathrm{in}} = 1$ $M_\odot$ yr$^{-1}$ as a fiducial value, but below we will explore how robust our results are against variations in this choice.

Observations indicate that the velocity dispersion of the densest molecular gas near the galactic centre is of order tens of km s$^{-1}$ (e.g., \citealt{walsh11a, purcell12a}, and Table 1 of \citealt{kruijssen14b}\footnote{These authors generally report either the FWHM or the 1D velocity dispersion, while our $\sigma$ is the 3D velocity dispersion, which is larger by a factor of $\sqrt{3}$.}). The velocity dispersion of somewhat lower density gas is much less constrained, but observations elsewhere in the Galaxy \citep{walsh04a, andre07a, kirk07a, rosolowsky08a}, as well as theoretical expectations \citep[e.g.][]{padoan01a, offner08a, offner09b}, suggest that it should be larger. We adopt $\sigma_{\mathrm{in}} = 40$ km s$^{-1}$ as our fiducial velocity dispersion, but, as with the mass inflow rate, we explore below how varying $\sigma_{\mathrm{in}}$ changes our results. 

As initial conditions for our simulations, we choose to place a small amount of material into the computational domain. Specifically, we set the initial surface density to $\Sigma = 1$ $M_\odot$ pc$^{-2}$ and the initial velocity dispersion to $\sigma = 40$ km s$^{-1}$ everywhere. The results are quite insensitive to these values as long as the mass and energy of the initial gas is small enough that the external inflow rapidly swamps the mass and internal energy of the material that was present at the beginning. For our fiducial parameter choices, externally-input material dominates after $0.6$ Myr.

All the source code for the simulations described in this section, and for the analysis presented in the subsequent sections, is publicly available. The run and analysis scripts can be obtained from \url{https://bitbucket.org/krumholz/cmzdisk/} (hash b98da00), and \texttt{VADER} code is available at \url{https://bitbucket.org/krumholz/vader} (hash 5a96350).

\section{Results}
\label{sec:results}

\subsection{Fiducial Case}

\begin{figure}
\includegraphics[width=\columnwidth]{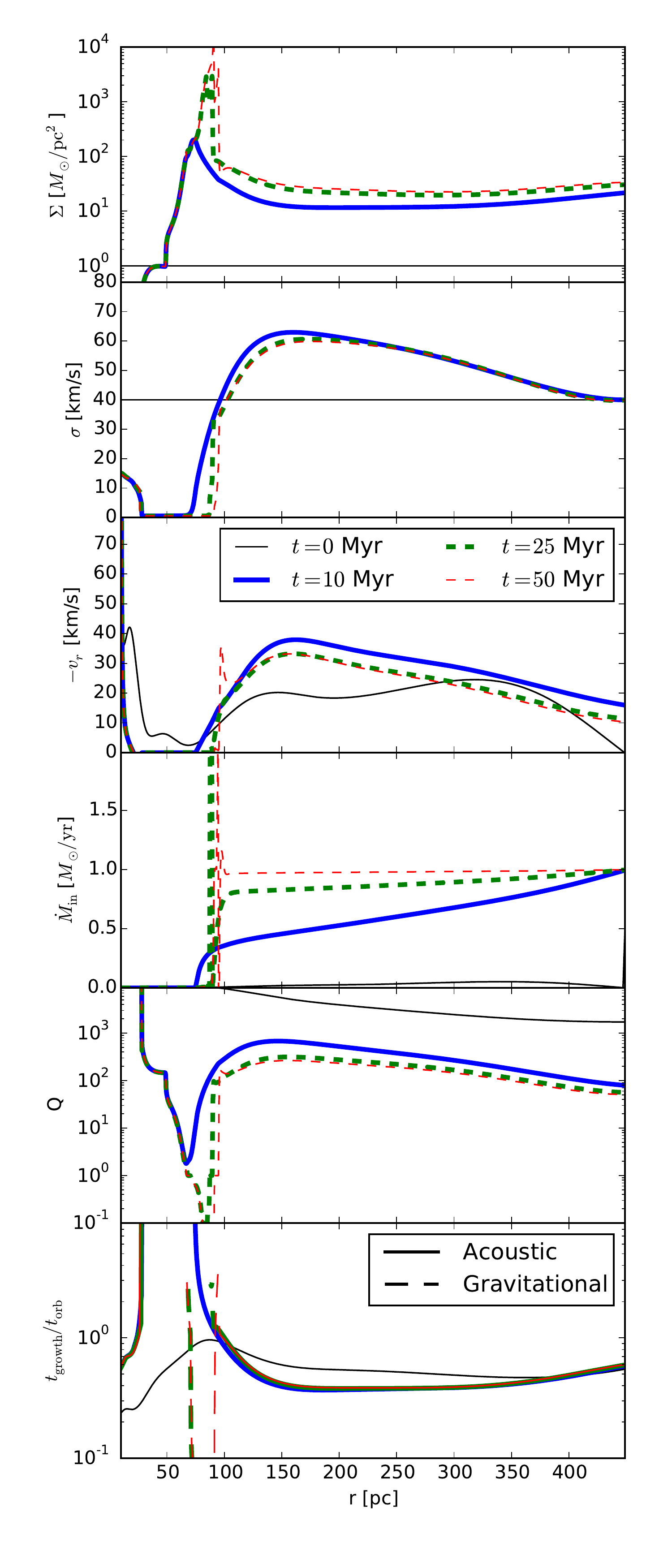}
\caption{
\label{fig:fiducial1}
Results of the simulation with the fiducial parameters. From top to bottom, the panels show the surface density $\Sigma$, velocity dispersion $\sigma$, inward radial velocity $-v_r$, instantaneous inflow rate $\dot{M}_{\mathrm{in}} = -2\pi r\Sigma v_r$, Toomre $Q$ parameter, and instability growth time normalized to the orbital time $t_{\mathrm{growth}}/t_{\mathrm{orb}}$. \red{In the top four panels, we show results at $t=0$ (\textit{thin black solid line}), $t=10$ Myr (\textit{thick blue solid line}), $t=25$ Myr (\textit{thick green dashed line}), and $t=50$ Myr (\textit{thin red dashed line}), as indicated in the legend.} For the instability growth time \red{in the bottom panel}, we show the value for the fastest-growing mode, with a solid line indicating that the fastest mode is acoustic and a dashed line indicating that it is gravitational. \red{Colors and line thicknesses are as in the panels above.}
}
\end{figure}

We first use the numerical method described in the previous section to simulate a fiducial case, for which the free parameters are as shown in \autoref{tab:fiducial}. \autoref{fig:fiducial1} shows the results of this simulation over 50 Myr. Qualitatively, we see that mass enters the computational domain from the outer edge and propagates inward through the disc as time passes. At early times the inflow rate is highest at the outer edge of the disc (where it is forced by our boundary conditions to be $1$ $M_\odot$ yr$^{-1}$), and then decreases inward. The velocity dispersion rises as the gravitational potential energy of this incoming material is converted into turbulent motion, at a rate that is initially too high for the decay of turbulence to counter it. This increase in velocity dispersion keeps $Q$ high at radii above $\sim 100$ pc. By $\sim 30$ Myr of evolution the disc outside $\sim 100$ pc has settled into a steady state whereby the inflow rate produced by acoustic instability is the same at all radii, and the velocity dispersion is kept constant by a balance between turbulent dissipation and the conversion of gravitational to turbulent energy as mass accretes down the potential well. In this equilibrium, $Q$ remains of order 100, so that the gas is extremely stable against collapse or star formation.

\begin{table}
\centering
 \begin{minipage}{80mm}
  \caption{Parameter values for the fiducial case. \label{tab:fiducial}}
\begin{tabular}{@{}lcc@{}}
\hline
Parameter & Value & Meaning \\
\hline
$\alpha_0$ & 1.0 & Dimensionless viscosity \\
$\dot{M}_{\mathrm{in}}$ & 1.0 $M_\odot$ yr$^{-1}$ & External accretion rate \\
$\sigma_{\mathrm{in}}$ & 40 km s$^{-1}$ & Incoming 3D velocity dispersion \\
$f_{\mathrm{shape}}$ & 2.5 & Potential shape parameter\\
$\sigma_{\mathrm{th}}$ & 0.5 km s$^{-1}$ & Thermal velocity dispersion \\
$\eta$ & 1.5 & Turbulence decay rate \\
\hline
\end{tabular}
\end{minipage}
\end{table}

\begin{figure}
\includegraphics[width=\columnwidth]{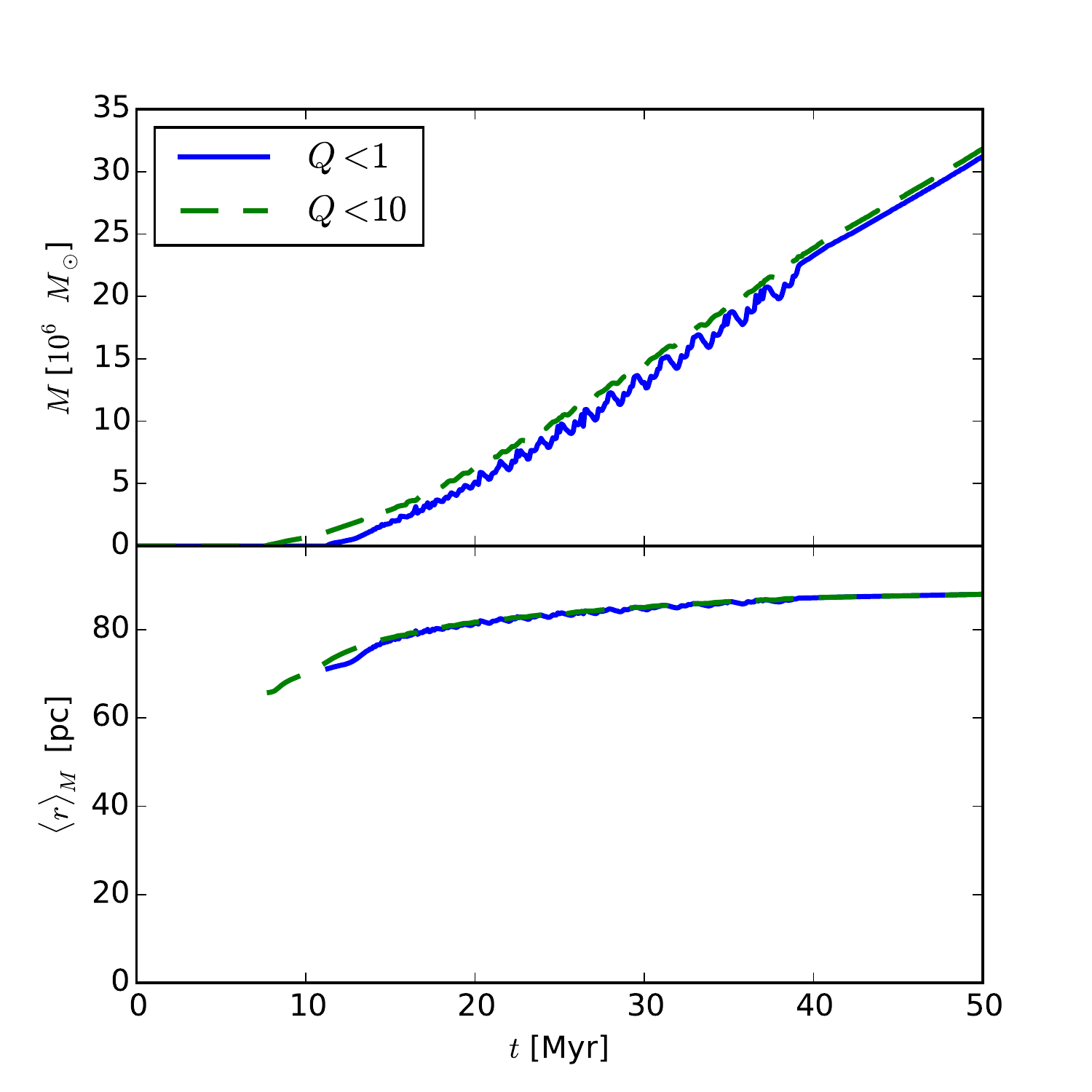}
\caption{
\label{fig:fiducial2}
Mass in the regions with $Q<1$ and $Q<10$ (\textit{top panel, \red{solid blue and dashed green lines, respectively}}), and mass-weighted mean radius of this mass (\textit{bottom panel}) versus time in the fiducial simulation. The mean radius is plotted only at times when the mass is non-zero.
}
\end{figure}

The results change qualitatively inside $\sim 100$ pc, as the wave of incoming material stops propagating inward and instead piles up in a ring. The inward mass flow rate also drops sharply as acoustic instability shuts off. At later times gravitational instability takes over, but this does not prevent a large pileup of gas. Within the accumulating ring of material, the velocity dispersion begins to drop, reaching the thermal floor value of $0.5$ km s$^{-1}$. As a result the $Q$ value in the ring begins to drop, first falling below unity at some point within the disc roughly $10-15$ Myr after the simulation starts. In \autoref{fig:fiducial2} we show the total amounts of mass for which $Q<1$ and $Q<10$, and the mass-weighted mean radius of this gas, as a function of time.\footnote{This computation is somewhat subtle, because the regions in which we are interested are, at early times, just a few computational cells wide. To suppress discreteness noise, we compute the mass in the unstable region via the following procedure. We take the simulation output at each time and compute $Q$ on our computational grid. We then construct \citet{akima70a} spline approximations to $\log r$ versus $\log \Sigma$ and $\log r$ versus $\log Q$ at each time step. We use the splines to generate finer grids of 32,768 points, and integrate the mass in the $Q<1$ and $Q<10$ regions on that higher resolution grid. Note that the small oscillations seen in mass versus time shown for $Q<1$ in \autoref{fig:fiducial2} are \textit{not} a result of this procedure, and are robust against changes in it. Instead, they are a real feature of the simulation. Oscillations occur because as $Q$ approaches unity, gravitational instability turns on as a transport mechanism. This leads parts of the disc to undergo a cycle in which regions where $Q<1$ have active mass transport that raises the velocity dispersion and drives $Q$ above unity so that transport stops. At that point, the turbulence begins to decay, so $Q$ eventually drops back below unity, turning on transport and starting the cycle again.} By $\sim 50$ Myr, several times $10^7$ $M_\odot$ of material has accumulated in the unstable region. Thus a mass comparable to that of the observed ring-like stream in the CMZ \citep{molinari11a} can accumulate in an unstable region after a few tens of Myr.

We can understand the outcome of the simulations by examining the rotation curve shown in \autoref{fig:rotcurve}. The region where the incoming material stalls is precisely where the rotation curve turns over from close to flat, as it is in the bulk of the Galaxy, to close to solid body, as it is near the Galactic Centre. In this near-solid body region, there is little shear, as shown by the dimensionless shear $1-\beta$. This suppresses transport in two ways. First, since acoustic instabilities are driven by the presence of shear, the approach to solid body rotation weakens the acoustic instability and thus lowers $\alpha$. Second, recalling \autoref{eq:torque}, we see that even for fixed $\alpha$ the torque is reduced in low-shear regions. This is physically what we should expect: for any plausible turbulent transport mechanism, the rate of angular momentum transport should be proportional to the rate of shear between adjacent rings. If there is no shear, transport mechanisms will shut down.

A third, less obvious effect of the low-shear region in on the energy balance of the disc. Energy transport in the disc occurs through two channels or roughly equal importance. Material advecting inward converts potential energy into kinetic energy as it does so, raising the local velocity dispersion, and torques within the disc transport kinetic energy outward from faster-moving material near the centre to slower-moving material further out. Both of these processes also tend to shut down in regions of low shear. As a result, the rate at which the velocity dispersion of material is being driven up by transport processes reaches a minimum in the near-solid body region. Since the rate of turbulent dissipation does not diminish there, this causes the local velocity dispersion to diminish, which in turn reduces the rate of turbulent transport to diminish even further.

As a result of these runaway effects, the low-shear region acts like a barrier that stops transport of material inward and energy outward, leading to a buildup of a low velocity dispersion, high surface density ring. The accumulation continues until gravitational instability sets in, at which point it seems likely that star formation will ensue and that star formation feedback will increase the velocity dispersion and possibly eject material entirely. The time required to accumulate enough mass to reach gravitational instability for the fiducial parameters is $\sim 15$ Myr.

\subsection{Dependence on Free Parameters}
\label{sec:free}

We next investigate to what extent the results we have obtained for the fiducial case are generic, and to what extent they depend on our poorly-known parameters. We summarize the parameters we vary, and the values we try, in \autoref{tab:param}. We vary the parameters one at a time, leaving all others fixed to the fiducial values indicated in \autoref{tab:fiducial}. In all cases we use the same initial conditions and computational grid, and run for 50 Myr. We do not vary the dissipation rate coefficient $\eta$, partly because it is reasonably well-calibrated from simulations, and partly because it is redundant with $f_{\mathrm{shape}}$. We also do not show the results of varying the thermal velocity dispersion $\sigma_{\mathrm{th}}$, because for plausible values of this parameter its effects are negligible.

\begin{table}
\centering
 \begin{minipage}{80mm}
  \caption{Variations in parameter values. \label{tab:param}}
\begin{tabular}{@{}lcc@{}}
\hline
Parameter & Values & Meaning \\
\hline
$\alpha_0$ & 0.1, 0.5, \textbf{1.0}, 2.0 & Viscosity \\
$\dot{M}_{\mathrm{in}}$ & $10^{-0.5}$, \textbf{1}, $10^{0.5}$ $M_\odot$ yr$^{-1}$ & Accretion rate \\
$\sigma_{\mathrm{in}}$ & 20, 30, \textbf{40}, 80 km s$^{-1}$ & Velocity dispersion \\
$f_{\mathrm{shape}}$ & 1.0, \textbf{2.0}, 4.0 & Potential shape\\
\hline
\end{tabular}
\medskip
Fiducial values are indicated in bold.
\end{minipage}
\end{table}

\begin{figure}
\includegraphics[width=\columnwidth]{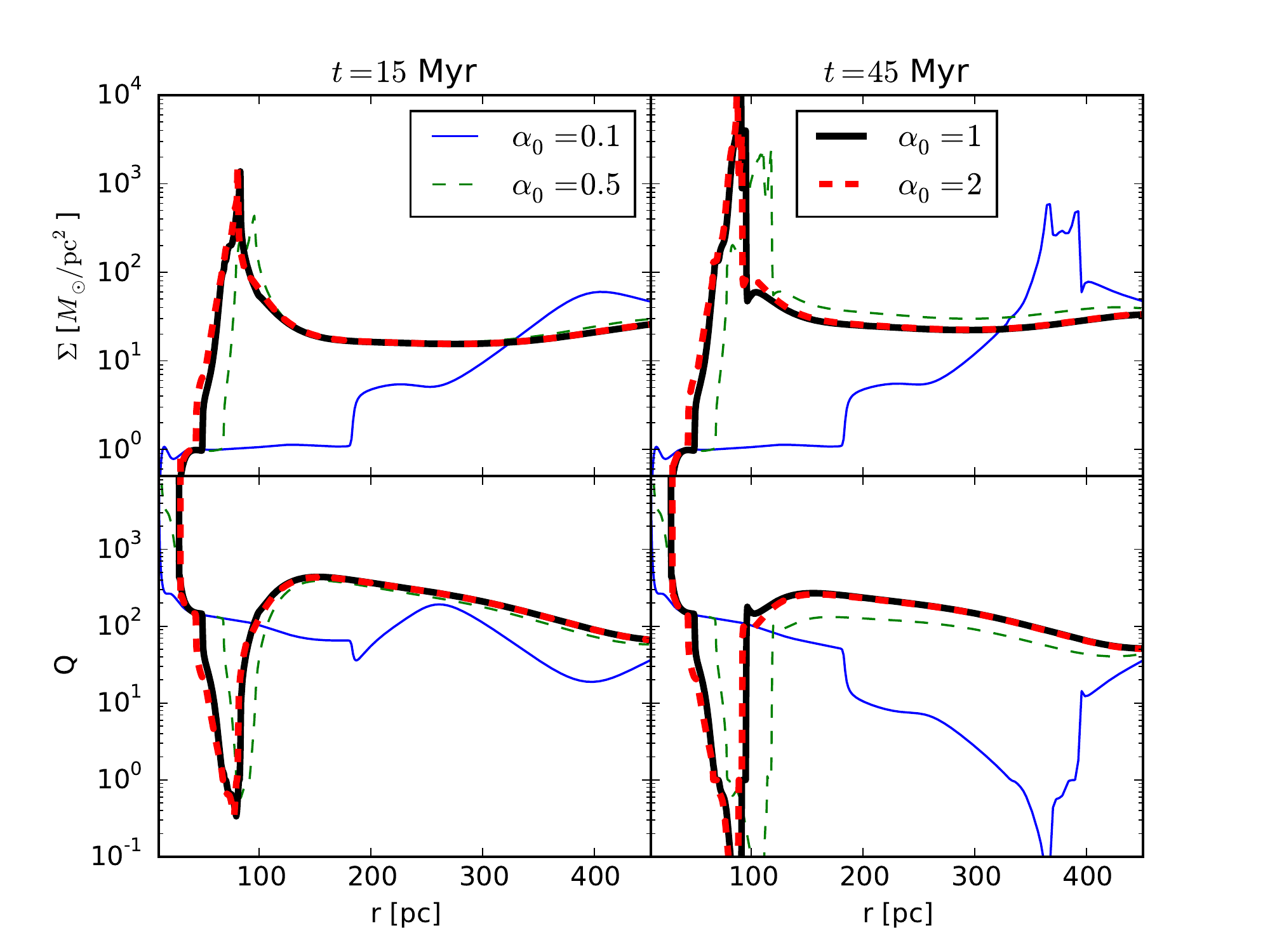}
\caption{
\label{fig:alphastudy1}
Results of the simulations with all parameters set to their fiducial values (\autoref{tab:fiducial}) except $\alpha_0$, which varies as indicated in the legend. The top row shows the surface density $\Sigma$ and the bottom shows $Q$; the left column shows results at $t=15$ Myr, and the right at $t=45$ Myr. \red{The fiducial case is shown as the solid thick black line, in this and all subsequent figures in this section.}
}
\end{figure}

\begin{figure}
\includegraphics[width=\columnwidth]{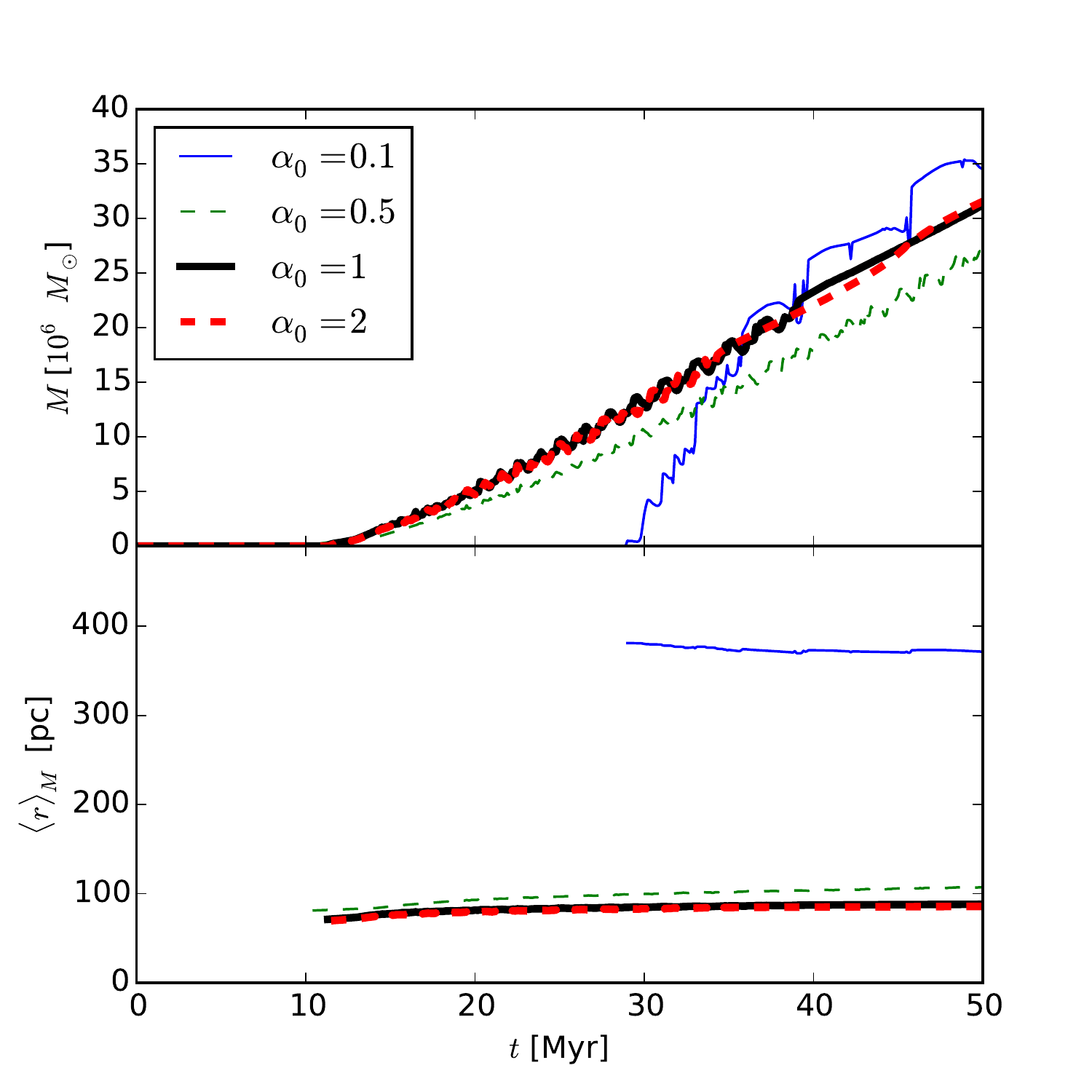}
\caption{
\label{fig:alphastudy2}
Same as \autoref{fig:fiducial2}, except that we only show the mass and radius of the material with $Q<1$ (not $Q<10$), and we show results for runs with varying values of $\alpha_0$ as indicated in the legend. The line for $\alpha_0 = 1$ is the same as in \autoref{fig:fiducial2}.
}
\end{figure}

We first vary $\alpha_0$, which controls the normalization for the rate of angular momentum transport. Figures \ref{fig:alphastudy1} and \ref{fig:alphastudy2} show how the results change as we vary $\alpha_0$ from its fiducial value of $1.0$ in the range $0.1 - 2.0$. We see that the results for all values of $\alpha_0 \geq 0.5$ are qualitatively the same as in the fiducial case. On the other hand, for $\alpha_0 = 0.1$ we obtain a qualitatively different result. In this case the angular momentum transport provided by the acoustic instability is insufficiently rapid to transfer mass inward at the rate of $1$ $M_\odot$ yr$^{-1}$ at which it enters the computational domain. As a result, the gas stagnates and collapses to form a gravitationally-unstable ring at a radius of $350-400$ pc rather than $\sim 100$ pc as in the other cases. Collapse is also significantly delayed relative to the fiducial case, taking $\sim 30$ Myr to evolve to a gravitationally-unstable state rather than $\sim 15$ Myr.

\begin{figure}
\includegraphics[width=\columnwidth]{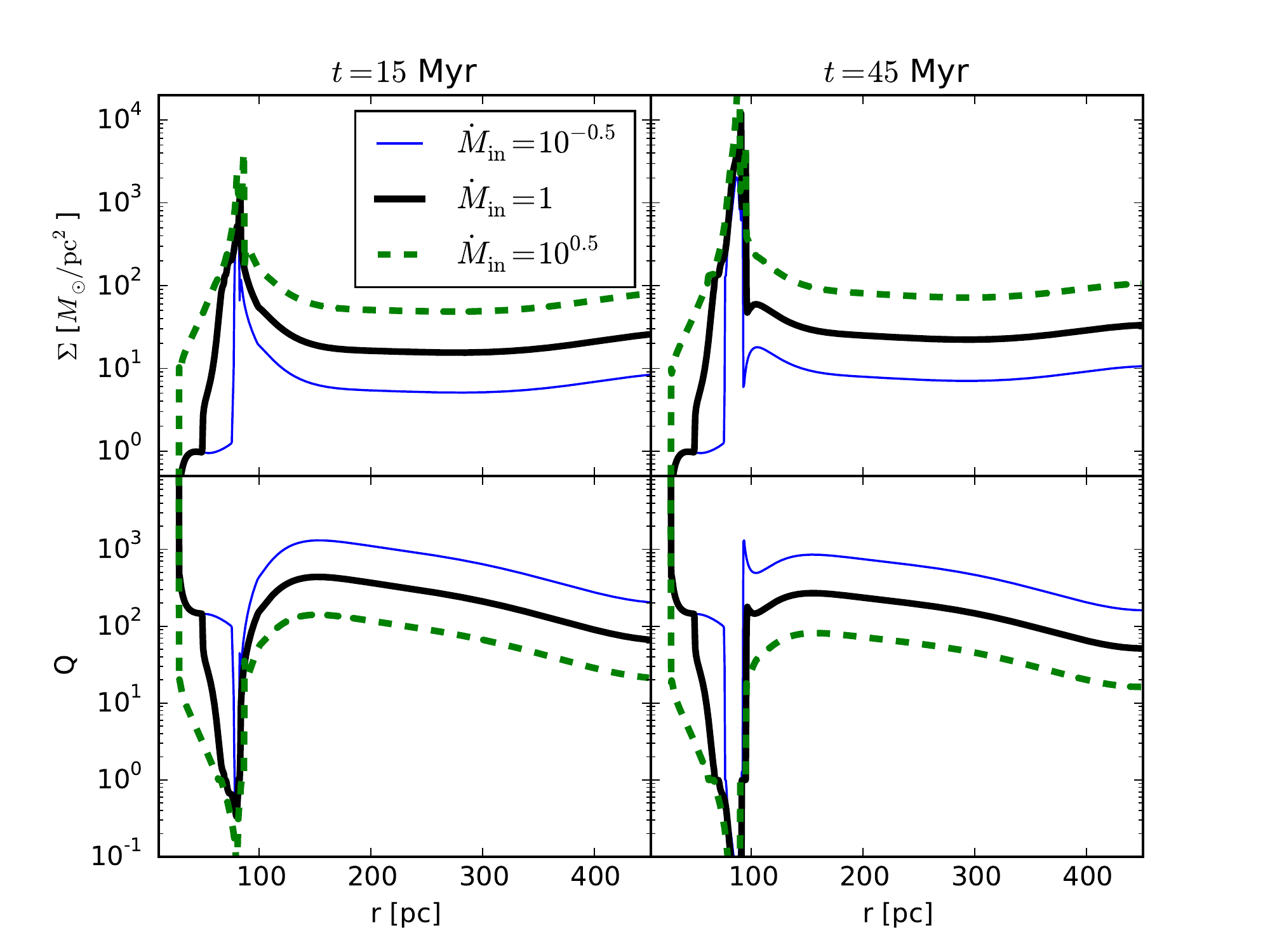}
\caption{
\label{fig:mdotstudy1}
Same as \autoref{fig:alphastudy1}, but with all parameters set to their fiducial values (\autoref{tab:fiducial}) except $\dot{M}_{\mathrm{in}}$, which varies as indicated in the legend (in units of $M_\odot$ yr$^{-1}$).
}
\end{figure}

\begin{figure}
\includegraphics[width=\columnwidth]{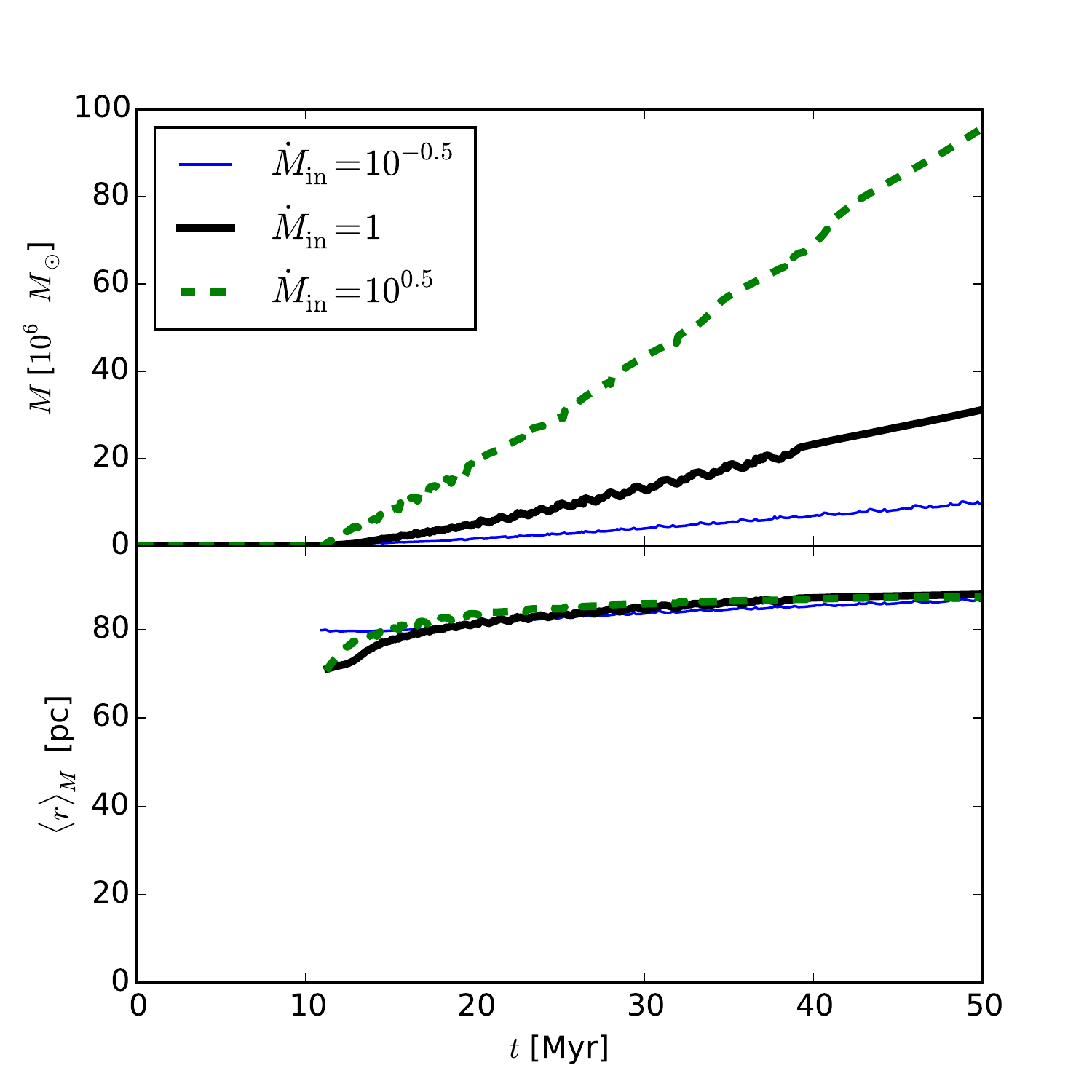}
\caption{
\label{fig:mdotstudy2}
Same as \autoref{fig:alphastudy2}, but for runs with varying values of $\dot{M}_{\mathrm{in}}$ as indicated in the legend (in units of $M_\odot$ yr$^{-1}$. The line for $\dot{M}_{\mathrm{in}} = 1$ is the same as in \autoref{fig:fiducial2}.
}
\end{figure}

We next try varying the inflow rate $\dot{M}_{\mathrm{in}}$, from $10^{-0.5} - 10^{0.5}$ $M_\odot$ yr$^{-1}$. Figures \ref{fig:mdotstudy1} and \ref{fig:mdotstudy2} show the results. Clearly varying the rate at which matter enters the CMZ from larger radii in the disc affects the overall timescale of the evolution, but mostly in a way such that the behavior remains self-similar. A higher accretion rate simply produces a larger overall surface density at fixed time, and a proportionately larger mass in the gravitationally unstable region, but the shape of surface density and Toomre $Q$ versus radius are unchanged. Similarly, in all three cases the location of the gravitationally unstable ring is roughly the same, and rate at which the mass in the gravitationally-unstable region grows eventually asymptotes to match the rate at which matter enters the computational domain. The formation of an unstable region is clearly robust against changes in the accretion rate.

\begin{figure}
\includegraphics[width=\columnwidth]{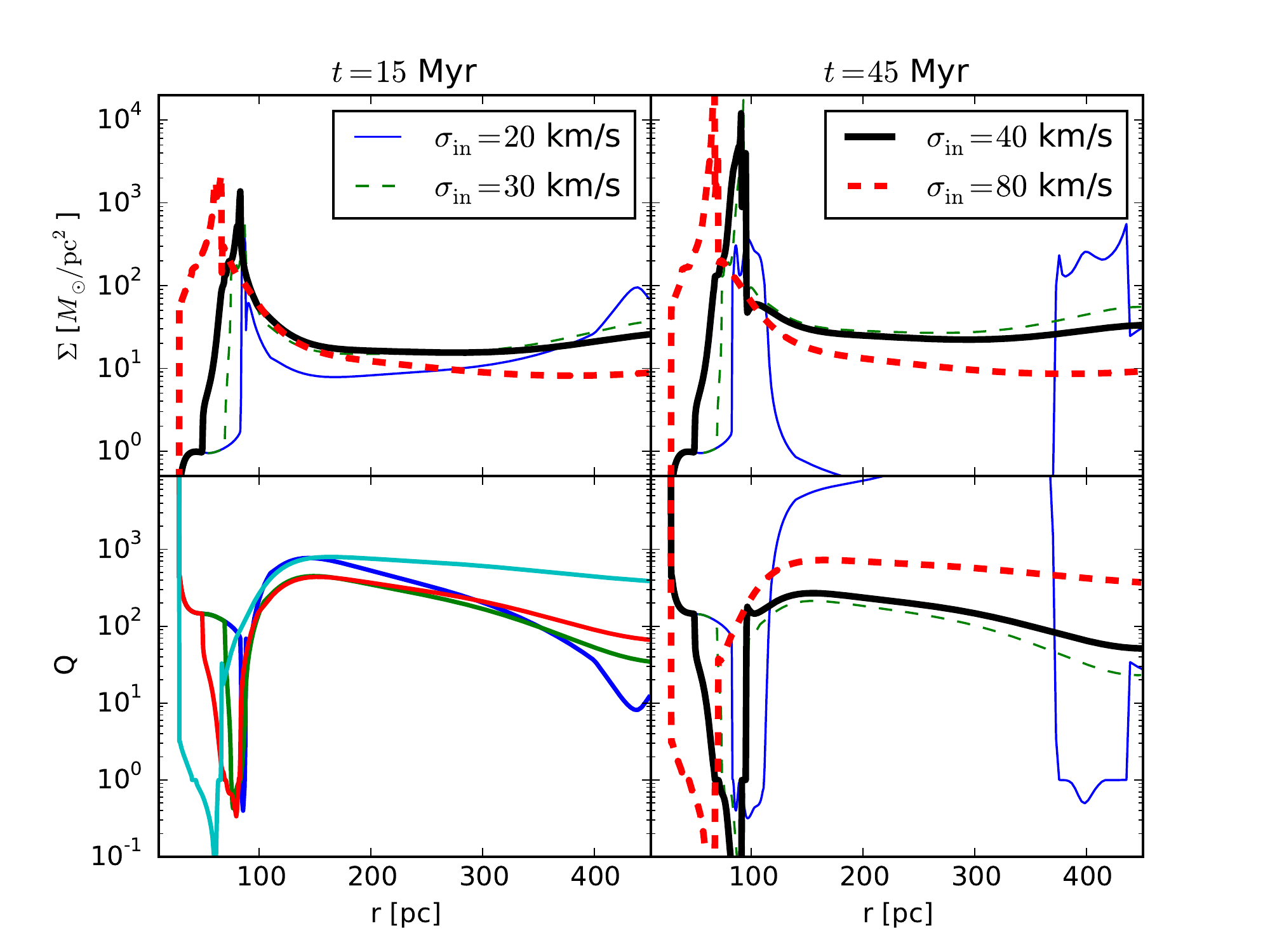}
\caption{
\label{fig:sigmastudy1}
Same as \autoref{fig:alphastudy1}, but with all parameters set to their fiducial values (\autoref{tab:fiducial}) except $\sigma_{\mathrm{in}}$, which varies as indicated in the legend.
}
\end{figure}

\begin{figure}
\includegraphics[width=\columnwidth]{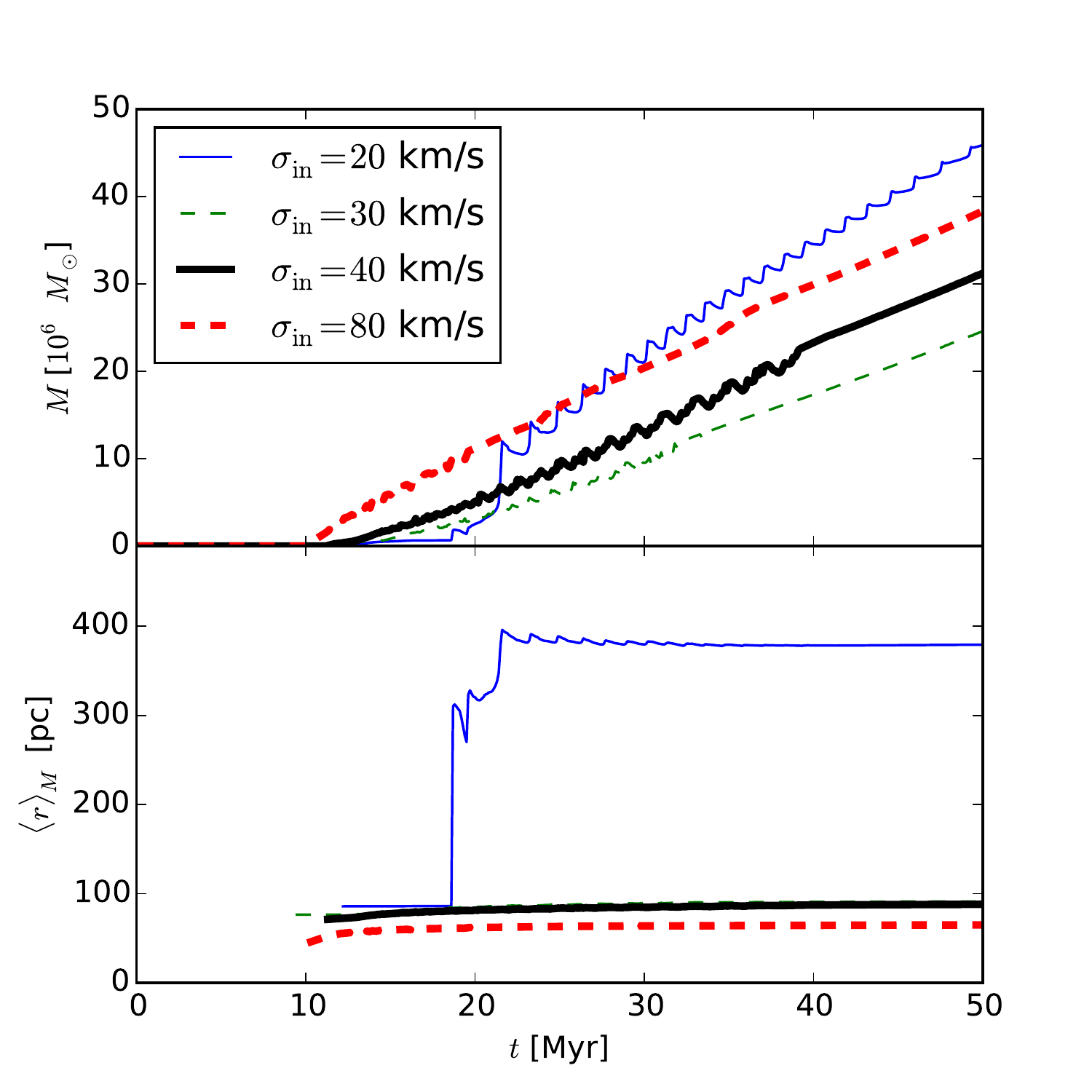}
\caption{
\label{fig:sigmastudy2}
Same as \autoref{fig:alphastudy2}, but for runs with varying values of $\sigma_{\mathrm{in}}$ as indicated in the legend. The line for $\sigma = 40$ km s$^{-1}$ is the same as in \autoref{fig:fiducial2}.
}
\end{figure}

Our third variation is in $\sigma_{\mathrm{in}}$, the velocity dispersion of the material entering the computational domain. This affects the rate of transport near the outer computational domain boundary, before the material advects inward far enough for its velocity dispersion to set by the balance between transport and dissipation. Figures \ref{fig:sigmastudy1} and \ref{fig:sigmastudy2} show the results of simulations using $\sigma_{\mathrm{in}} = 20$, 30, 40, and 80 km s$^{-1}$, roughly spanning the plausible observed range for CMZ material. The results for 30, 40 and 80 km s$^{-1}$ are clearly very similar qualitatively. For 20 km s$^{-1}$, the results are similar at times before $\sim 20$ Myr, but after that point the low velocity dispersion of the material entering the computational domain leads to lower rates of transport at large radii, and this in turn causes mass to build up at large radii. Eventually the gas builds up to form a second gravitationally-unstable region near $\sim 400$ pc, in addition to the first one formed at $\sim 100$ pc. Exploration of different values of $\sigma_{\mathrm{in}}$ shows that this secondary unstable region appears for values of $\sigma_{\mathrm{in}}$ below roughly $20-25$ km s$^{-1}$.

\begin{figure}
\includegraphics[width=\columnwidth]{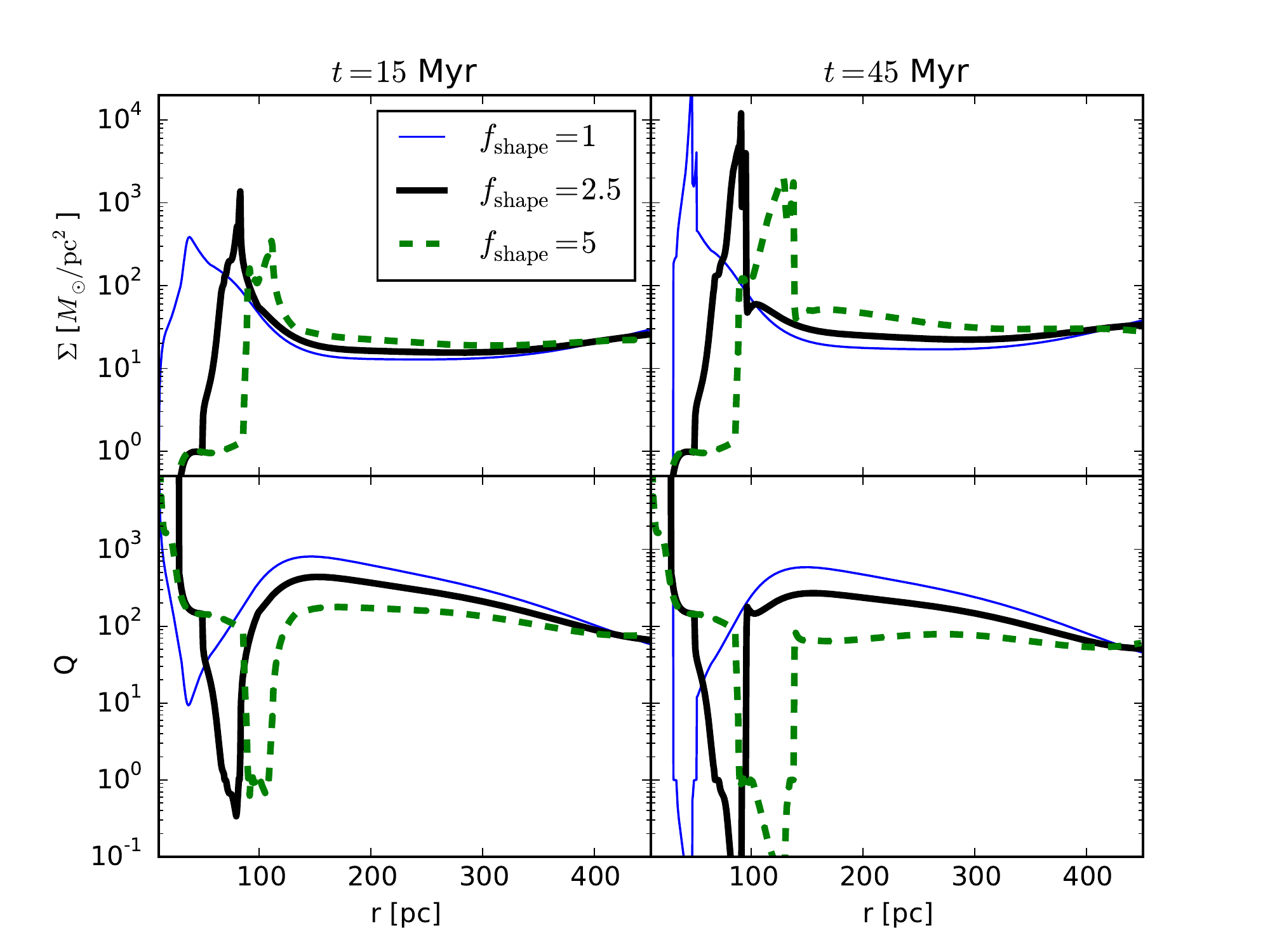}
\caption{
\label{fig:fshapestudy1}
Same as \autoref{fig:alphastudy1}, but with all parameters set to their fiducial values (\autoref{tab:fiducial}) except $f_{\mathrm{shape}}$, which varies as indicated in the legend.
}
\end{figure}

\begin{figure}
\includegraphics[width=\columnwidth]{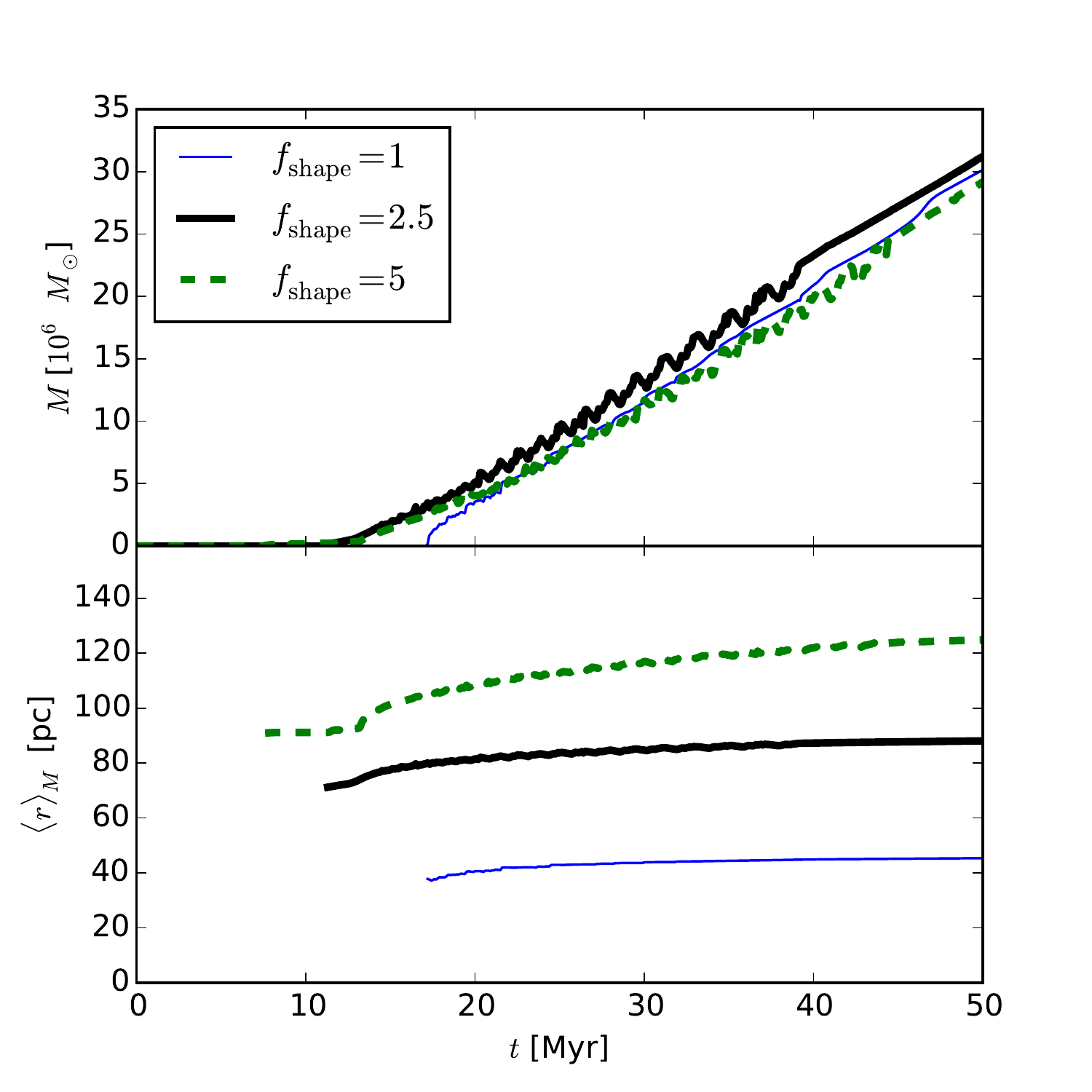}
\caption{
\label{fig:fshapestudy2}
Same as \autoref{fig:alphastudy2}, but for runs with varying values of $f_{\mathrm{shape}}$ as indicated in the legend. The line for $f_{\mathrm{shape}} = 2$ is the same as in \autoref{fig:fiducial2}.
}
\end{figure}

Finally, we vary the parameter $f_{\mathrm{shape}}$ that describes how flattened the stellar potential is, and thus how large the gas scale height is at fixed gas surface density and velocity dispersion. Since the turbulent dissipation time is set by the scale-height crossing time, this choice affects the rate at which the turbulence decays. \autoref{fig:fshapestudy1} and \autoref{fig:fshapestudy2} show the results of this experiment. Clearly the value of $f_{\mathrm{shape}}$ changes the precise location of the ring of gas buildup and gravitational instability, but does not alter the qualitative result that such a ring forms on timescales of $\sim 10-20$ Myr, or the default distribution of the gas approaching that ring.

To summarize our findings: gas entering the CMZ tends to hang up in its inward flow and develop a gravitationally-unstable region at $\sim 100$ pc, where the rotation curve has a region of minimal shear as it approach solid body. This result is very robust against variations in the mass accretion rate and the shape of the stellar potential. The former only acts as a scaling parameter, while the latter changes the precise location of the unstable region but not its general characteristics. This tendency is also robust against increases in the velocity dispersion of the incoming gas and the rate of angular momentum transport produced by acoustic instabilities, and to some level of decrease in these parameters as well. However, if the input velocity dispersion is too low, or the rate of angular momentum transport by acoustic instability too small, the gas collapses into gravitational instability near the edge of our simulation domain, either instead of or in addition to in the location of minimal shear at $\sim 100$ pc. Despite this, we can be reasonably confident that, over a broad range of parameters, buildup of a $\sim 100$ pc unstable ring should occur on timescales of tens of Myr.

\section{Discussion}
\label{sec:discussion}

\subsection{Comparison to the Structure of the Galactic CMZ} \label{sec:applycmz1}
The results presented in this paper show quantitatively how gravitational collapse is inhibited within the ILR resonance of galactic centres (and of the CMZ of the Milky Way in particular). This is caused by efficient angular momentum transport in acoustic instabilities, which maintains the turbulent velocity dispersion at a consistently high level ($\sigma=\sqrt{3}\sigma_{\rm 1D}>40~\kms$). In \citet{kruijssen14b}, it was proposed that the observed high turbulent pressure ($\sigma\sim40~\kms$ and $P/k\sim10^8~{\rm K}~\cmc$, see e.g., \citealt{bally88a}) and the correspondingly high Toomre stability parameter ($Q\sim10$) of the CMZ gas are responsible for the fact that the star formation rate in the CMZ is observed to be an order of magnitude lower than expected from empirical scaling relations and star formation theory \citep{longmore13a}. It remained an open question whether the high turbulent pressure is driven by stellar feedback, the gas inflow along the bar, or acoustic instabilities. In the present paper, we find that acoustic instabilities naturally lead to the observed, high velocity dispersions. Our model therefore provides a key missing element in explaining the currently low star formation rate in the CMZ.

\begin{figure*}
\includegraphics[width=\hsize]{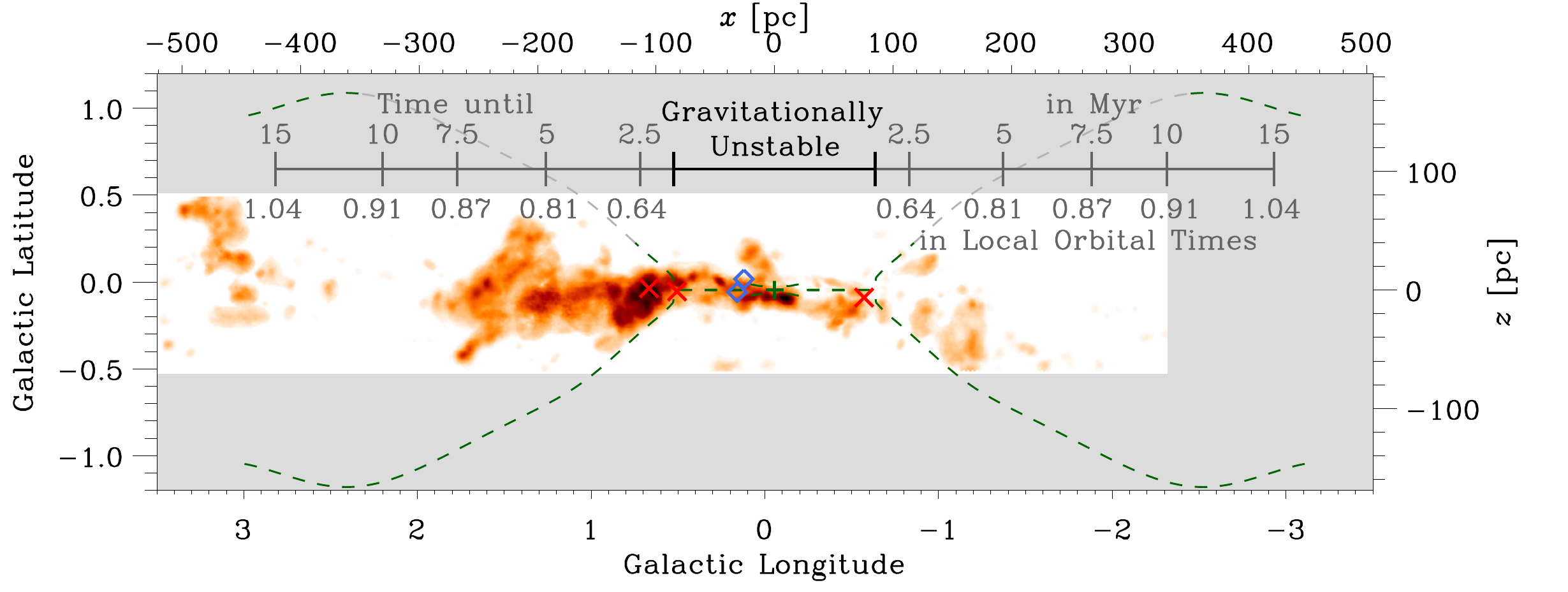}
\caption{
\label{fig:cmzmap}
Column density map of the dense ($n>{\rm several}~10^3~\cmc$) gas in the Galactic CMZ as traced by NH$_3(1,1)$ \citep{walsh11a,purcell12a}, with symbols indicating Sgr~A$^*$ (\textit{green plus}), star-forming clouds (\textit{red crosses}; from left to right these are Sgr~B2, Sgr~B1, and Sgr~C), and young stellar clusters (\textit{blue diamonds}; from left to right these are the Quintuplet and Arches clusters). The overlay provides several of our model predictions at $t=17.5~\myr$ in the plane perpendicular to the line of sight at the distance of Sgr~A$^*$. The dashed lines show the scale height profile predicted for {\it all} gas (i.e.~not just the dense gas shown here), whereas the scale bars indicate the time necessary for the gas at each longitude to reach the gravitationally unstable region (black line), in units of $\myr$ (top) as well as normalised to the {\it local} orbital time at each radius (bottom). As the gas is transported inwards, the remaining number of orbital times decreases more slowly than the remaining absolute time, illustrating a decrease of transport rate. As a result, the entire migration process proceeds over several orbital revolutions and the gas piles up just outside the gravitationally unstable region.
}
\end{figure*}

Next to providing an accurate description of the gas in the CMZ on large scales, our model also reproduces the presence of a gravitationally unstable, ring-like stream of gas \citep{molinari11a,kruijssen15a}. As gas loses angular momentum and flows towards the Galactic centre, it accumulates at a radius of $\sim100~\pc$, where the shear has a local minimum. This reduces the angular momentum transport rate and the driving of turbulence by acoustic instabilities. Consequently, the gas accumulates and the surface density increases, until after $10$--$15~\myr$ a gravitational instability develops. For the fiducial parameter set, the model predicts a peak surface density of $\Sigma>10^3~\msun~\pc^{-2}$ and a minimum Toomre $Q=0.3$--$1.0$ at $t=15~\myr$, briefly after the gravitational instability has set in. These numbers are in remarkable agreement with the observed properties of the gas stream (cf.~Table~1 of \citealt{kruijssen14b}).\footnote{In our model, the typical values of $\Sigma$ and $Q$ within the gravitationally unstable radius interval are primarily set by $\dot{M}$ and $\sigma_{\rm in}$ (see \autoref{sec:free}). For good agreement with the observations, the model requires $\dot{M}\sim1~\msun~\yr^{-1}$ and $\sigma_{\rm in}>30~\kms$. The former is consistent with other theoretical constraints \citep[e.g.,][]{crocker12a}, whereas the latter of these is consistent with the observed velocity dispersions mentioned above.}

To make a more detailed comparison, we focus on the state of our fiducial simulation at 17.5 Myr of evolution, which we compare to the observed CMZ in \autoref{fig:cmzmap}; we select this time slice because it has depletion times in good agreement with those observed in the present-day CMZ (see \autoref{sec:applycmz2} for details). The figure illustrates that the gravitationally unstable region matches the radii where currently most of the dense gas, star-forming regions, and young stellar clusters in the CMZ reside \citep[e.g.,][]{molinari11a,longmore13b}. In our model, this region covers a closed circular ring by definition, because we assume axisymmetry on the sub-kpc scales under consideration. However, the gas itself is expected to follow a possibly eccentric, stream-like structure within this unstable radius interval, which must be open-ended due to the extended nature of the stellar mass distribution. The radial range across which the gravitational instability takes place in our parameter survey is $60\la R/\pc\la120$, which indeed matches the radii covered by the eccentric orbital model for the CMZ gas stream by \citet{kruijssen15a}.

At radii beyond the gravitationally unstable region (i.e.~$150<R/\pc<500$ or $1^\circ\la|l|\la3.5^\circ$ in projection), dense clouds may exist, but in our model they have high velocity dispersions and are supervirial due to acoustic instabilities ($\alpha_{\rm vir}\sim P_{\rm turb}/P_{\rm grav}=\sigma^2/\pi G H_{\rm g}\Sigma=10$--$100$). Therefore, their global collapse is inhibited and the little star formation activity they are expected to have will result from stochastic turbulent motion (cf.~Bania's Clump, \citealt{bally10a}). For the CMZ model presented here, star formation theories \citep[e.g.,][]{krumholz05c, padoan12a, federrath12a, hennebelle13a} predict low star formation efficiencies per free-fall time of $\epsilon_{\rm ff}\sim0.001$ outside the gravitationally unstable region. \red{This is particularly true if the turbulence is primarily solenoidal \citep{federrath12a}, as might be expected since it is ultimately driven by shear.}

As illustrated in \autoref{fig:cmzmap}, the high velocity dispersions result in an increase of the gas scale height with radius. Depending on their Galactic longitudes, it takes these highly-turbulent clouds anywhere between $2$--$15~\myr$ to migrate inwards to the gravitationally unstable region at $R\sim100~\pc$ (cf.~the third panel of \autoref{fig:fiducial1}, which shows inward radial velocities of $-v_{\rm r}=20$--$30~\kms$), corresponding to $\sim2$ full orbital revolutions. This estimate assumes that the clouds all reside at the same distance -- for non-zero displacements along the line of sight, the migration time-scales are even longer. Direct-infall models for the gas inflow onto the CMZ (e.g., those in which the $x_1$ and $x_2$ orbits play a major role) predict much shorter migration time-scales of $0.5$--$5~\myr$ \citep[e.g.][]{sofue95a,bally10a}. Comparing observed cloud properties as a function of longitude with the expected evolutionary time-scales may therefore be a way of discriminating between direct-infall models and the dynamical evolution model presented here. Other tests of the model can be performed by comparing the radial profiles of e.g., the gas surface density and velocity dispersion from \autoref{fig:fiducial1} to observed maps of the CMZ.

As a result of the inward transport of the gas and the corresponding increase of the gas pressure towards lower Galactic longitudes, our model predicts that the dense gas fraction increases strongly towards the Galactic Centre. To quantify this prediction, we consider gas near the disc midplane, where the mean density is $\rho = \Sigma/2H_g$. We then assume that the gas has a log-normal volume density PDF as expected for supersonically turbulent, isothermal media \citep[e.g.][]{Vazquez-Semadeni94a,padoan97a,krumholz05c}, and calculate the mass fraction above a certain minimum density.\footnote{\red{The density PDF can be altered somewhat depending on the ratio of solenoidal to compressive modes in the driving force \citep{federrath08a}, the strength of magnetic fields \citep{molina12a}, and deviations from isothermality \citep{federrath15a}, but for simplicity we perform this calculation using the results for an isothermal, non-magnetic medium with mixed solenoidal-compressive driving.}} The resulting dense gas fractions are listed at four different Galactic longitudes in \autoref{tab:dense}. Interestingly, the dense gas fractions change only slowly in the outer CMZ, where acoustic instabilities inhibit gravitational collapse. However, they increase rapidly as the gas becomes gravitationally unstable at $|l|<1^\circ$. Designating all gas with densities $n>10^4~\cmc$ as `dense', we find that the dense gas fraction increases from a few per cent in the outer CMZ to nearly unity in the 100-pc ring (as was also found by \citealt{longmore13a}). These fractions decrease strongly as the minimum density used to define `dense' gas increases. Interferometric observations of high-critical density gas tracers using ALMA or the ongoing SMA Legacy Survey of the CMZ (Keto et al.~in preparation; Battersby et al.~in preparation) are ideally suited to test these predictions.
\begin{table}
 \centering
  \begin{minipage}{80mm}
  \caption{Predicted dense gas fractions in the Galactic CMZ.}\label{tab:dense}
  \begin{tabular}{@{}l c c c@{}}
  \hline 
  Longitude & $f_{\rm dense,4}$ & $f_{\rm dense,5}$ & $f_{\rm dense,6}$ \\
   & $(>10^4~\cmc)$ & $(>10^5~\cmc)$ & $(>10^6~\cmc)$ \\
 \hline
$|l|=2$--$3^\circ$ & $0.03$ & $0.003$ & $0.0002$ \\
$|l|\sim1.5^\circ$ & $0.05$ & $0.006$ & $0.0005$ \\
$|l|\sim1^\circ$ & $0.08$ & $0.015$ & $0.0015$ \\
$|l|<0.7^\circ$ & $0.3$--$1.0$ & $0.04$--$0.2$ & $0.005$--$0.01$  \\
\hline
\end{tabular}
These values apply to our fiducial model (\autoref{fig:fiducial1}). While the overall trend of increasing dense gas fractions towards lower longitudes is a robust prediction of the model, the normalisation of that trend will depend on the adopted parameters (cf.~\autoref{sec:free}). As these parameters become better constrained by new observations, so will the dense gas fractions provided here.
\end{minipage}
\end{table}

\subsection{Time-evolution of the Galactic CMZ} \label{sec:applycmz2}
It is important to reiterate that our model predicts an episodic cycle for the gas content and star formation activity of galactic centres. The development of a gravitationally unstable region marks the beginning of the end: once the gas collapses and forms stars, the residual gas is expelled by feedback. The current gas mass of the 100-pc stream in the CMZ is $M\sim10^7~\msun$, with a density of $n>10^4~\cmc$ and a correspondingly short free-fall time of $t_{\rm ff}<0.34~\myr$, which is much shorter than the time required to grow the current gas reservoir ($t_{\rm acc}\sim M/\dot{M}_{\rm in}\sim10~\myr$ for our fiducial model). As a result, the gravitationally unstable gas reservoir is expected to be depleted by a combination of rapid star formation and feedback, after which the accretion cycle described by our model starts from the beginning.

Given the episodic nature of the system's evolution, a more detailed comparison between model and observations requires us to assess at which point along the cycle the CMZ currently resides. The present star formation rate in the CMZ is only $\sim0.05~\msun~\yr^{-1}$ \citep{longmore13a,koepferl15a} and the mass outflow rate is estimated at $\sim0.5~\msun~\yr^{-1}$ (corresponding to a mass loading factor of $\eta_{\rm ml}\sim10$, see e.g., \citealt{crocker12a}). The sum of both ($\dot{M}_{\rm out}\sim0.6~\msun~\yr^{-1}$) is lower than our fiducial mass inflow rate of $\dot{M}_{\rm in}\sim1~\msun~\yr^{-1}$, and also lower than the independent estimate by \citet{crocker12a}, who finds $2\sigma$ limits of $0.4<\dot{M}_{\rm in}/\msun~\yr^{-1}<1.8$. It is therefore most probable that the gas mass in the CMZ is presently increasing. This places the CMZ at the moment prior to the starburst, but after the gravitational instability has started to develop.

Comparing the current mass of the 100~pc gas stream ($M\sim10^7~\msun$) to the time-evolution of the unstable mass in our fiducial model (see \autoref{fig:fiducial2}), the CMZ appears to reside some $t\sim20~\myr$ after its last major starburst. However, this is an upper limit, as it assumes that (1) the entire gas reservoir needed to be regrown out to $R>400~\pc$, and (2) that all of the mass in the gas stream is already gravitationally unstable. In the more likely scenario that the previous starburst did not clear out any of the gas beyond the unstable region, the evolution towards gravitational instability during the first $\sim10~\myr$ is skipped, because accretion can resume without delay. The growth phase of the gravitational instability then remains, resulting in a time interval of $\sim10~\myr$ to accumulate the present-day gas mass of the 100~pc stream. If the previous starburst did not clear out all of the gas in the unstable region, this time interval will be further reduced. We therefore conclude that the CMZ is currently $5$--$10~\myr$ post-starburst.\footnote{The main observational relics of this previous starburst are likely the 8-kpc {\it Fermi} bubbles that extend from the CMZ \citep{su10a}. In combination with their spatial extent, the time since the starburst given here constrains the outflow velocity of the bubbles to be $v_{\rm out}=800$--$1600~\kms$, which is remarkably consistent with previous estimates from geometric considerations \citep[$v_{\rm out}\sim1000~\kms$,][]{carretti13a}.}

\begin{figure}
\includegraphics[width=\hsize]{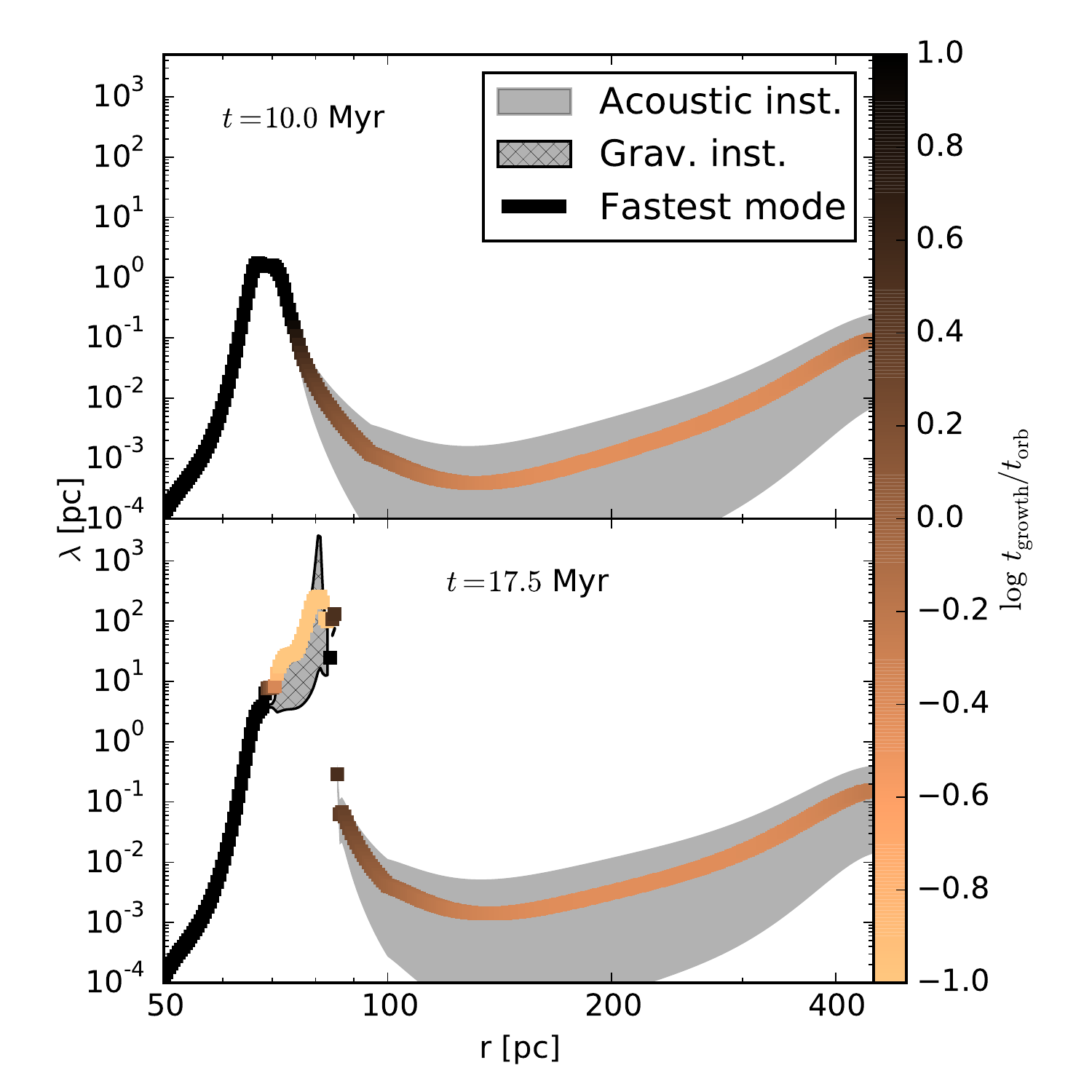}
\caption{
\label{fig:wavelength}
Unstable wavelengths for the acoustic (shaded area) and gravitational (hashed area) instabilities in the fiducial model, with the fastest-growing mode colour-coded by the growth time-scale in units of the orbital time (see colour bar). The top panel shows the radial profile prior to the onset of the gravitational instability at $t=10~\myr$, whereas the bottom panel shows the ``current-day" state at $t=17.5~\myr$.
}
\end{figure}

The obvious next question to ask is when the CMZ will enter the next starburst phase. To address this point, we consider our model at $t=17.5~\myr$, shortly after the initial formation of the gravitationally-unstable mass reservoir. If the properties of these gravitational instabilities match those observed in the CMZ, their growth time-scales provide insight in the CMZ's immediate evolution. \autoref{fig:wavelength} shows the length-scale of the acoustic and gravitational instabilities as well as the growth time-scale of the fastest-growing mode as a function of radius in the CMZ, at times $t=\{10,17.5\}~\myr$. For the acoustic instabilities, the figure generalises the initial result from \autoref{fig:tgrowth} to the evolved CMZ model, showing that the growth time-scale of these instabilities is typically less than an orbital time and that they develop rapidly enough to drive angular momentum transport and turbulence.\footnote{The characteristic length-scale of the instabilities is short, at $\lambda<0.01~\pc$. However, it is important to recall that this wavelength characterizes the radial size in the tight-winding approximation; that is, it is the characteristic width of the $m=2$ spiral features, and tells us nothing about the pitch angle of the spiral, other than the fact that it must be small since the \citet{montenegro99a} dispersion relation is derived in the limit where it is. Moreover, we emphasize that the fastest growing linear mode is not necessarily the dominant one in the non-linear regime, and thus the dominant mode that we would expect to observe. An obvious example is Rayleigh-Taylor instability, for which the growth rate in the linear regime scales as $\lambda^{1/2}$, so that small-scale modes are fastest, but both simulations and experiments show that long-wavelengths mode dominate in the non-linear regime \citep[e.g.,][]{dimonte04a}. In the absence of full simulations of the acoustic instability, we cannot reach any strong conclusions about which modes will dominate in real galaxies.}

The development of the gravitational instabilities around $R\sim100~\pc$ differs from the acoustic instabilities. \autoref{fig:wavelength} demonstrates that the growth time-scale is initially longer than an orbital time.\footnote{This supports the scenario put forward by \citet{longmore13b} and \citet{kruijssen15a} in which the collapse of clouds on the 100-pc stream is triggered by a tidal compression during pericentre passage. Because the gas on the stream initially requires several orbital revolutions to become gravitationally unstable, the recurrent tidal perturbations at pericentre are statistically likely to give the final nudge into collapse as the turbulent energy is gradually dissipating.} As the instability develops, the growth time-scale decreases substantially to $t_{\rm growth}<t_{\rm orb}$, and two modes with different wavelengths dominate:
\begin{enumerate}
\item At radii $R=80$--$100~\pc$, the instability length-scale is $\lambda\sim200~\pc$. This long-wavelength instability is expected to drive large-scale asymmetries in the distribution of the gas in the CMZ, as it covers a significant fraction of a circular orbit at these radii. This matches the well-known observation that most of the gas in the CMZ resides at positive Galactic longitudes (see \autoref{fig:cmzmap} and \citealt{bally10a}).
\item At radii $R=70$--$80~\pc$, the fastest-growing mode is a factor of several shorter, with $\lambda~\sim30~\pc$. This short-wavelength instability matches the estimated Jeans length in the 100-pc stream \citep[$\sim30~\pc$,][]{kruijssen15a}, as well as the observed typical separation length of the individual clouds condensing out of the stream and the wavelength of line-of-sight velocity oscillations (Henshaw et al.~in preparation).
\end{enumerate}
Given how well these instabilities match the observed density fluctuations in the CMZ, their growth time-scales may be taken as an indication of how rapidly the CMZ clouds can start their collapse towards the next starburst phase. 

Both of the above gravitationally unstable modes have $t_{\rm growth}<t_{\rm orb}\sim3.7~\myr$. Therefore, the extant condensations should start collapsing within the next few~Myr, and the precise time delay till the next starburst within the gravitationally unstable region will depend on the number of free-fall times needed to fundamentally change the ratio between gas mass and star formation rate. We use the observational estimate that star formation efficiencies of $\epsilon_{\rm sf}\sim0.1$ are reached before a star formation event disperses the residual gas \citep{lada03a, federrath13a}. The associated time-scale is $t_{\rm sf}=t_{\rm ff}\epsilon_{\rm sf}/\epsilon_{\rm ff}$, where $\epsilon_{\rm ff}$ is the star formation efficiency per free-fall time. In the Galactic disc, $\epsilon_{\rm ff}\sim0.01$ \citep{krumholz07e,krumholz12a, federrath13c}, but it is unclear if this holds in the CMZ as well.\red{\footnote{
\red{When globally-averaged surface densities are adopted, the CMZ is consistent with galactic star formation relations and models assuming $\epsilon_{\mathrm{ff}} \sim0.01$ \citep{yusef-zadeh09a, longmore13a, kruijssen14b, salim15a}, but this approach ignores persistent (and hence physical) substructure. Therefore, it does not directly constrain the true value of $\epsilon_{\mathrm{ff}}$ in the CMZ.}}} Using orbital modelling, \citet{kruijssen15a} estimate that the evolutionary time difference between the `Brick' (little star formation) to Sgr~B2 (high star formation activity) is $\Delta t\sim t_{\rm ff}\sim0.4~\myr$, whereas the star formation efficiency in Sgr~B2 is of the order a per cent \citep[assuming a few~$100~\msun$ per ultracompact H{\sc ii} region]{bally10a}. As a result, $\epsilon_{\rm ff}\sim0.01$ in collapsing CMZ clouds too. Combining the above numbers, we estimate that the next starburst in the gravitationally unstable region will be reached in $\sim5~\myr$ or 1--2 orbital revolutions.

\subsection{A Cartoon Model with Star Formation and Feedback} \label{sec:applycmz3}

We are now in a position to develop a toy model for the overall behaviour of star formation in central molecular zones, both those of the Milky Way and of similar galaxies, and how they evolve in the observational plane of star formation rate surface density versus gas surface density. Our goal here is not to precisely reproduce the properties of the Milky Way's CMZ. Instead, it is to develop a simple cartoon picture for how regions such as the CMZ evolve in time. To do so, we must extend our model with a simple treatment of star formation and feedback.

To this end, we modify our fiducial model by adding a term $-\dot{\Sigma}_*$ on the right-hand side of \autoref{eq:masscons}. We compute the star formation rate as
\begin{equation}
\dot{\Sigma}_* = f_{\mathrm{cl}} \epsilon_{\mathrm{ff}} \frac{\Sigma}{t_{\mathrm{ff}}}
\end{equation}
where $f_{\mathrm{cl}}$ is the fraction of gas in dense clouds, $\epsilon_{\mathrm{ff}}$ is the star formation rate per free-fall time in those clouds, $t_{\mathrm{ff}}$ is the free-fall time. We compute these quantities as $f_{\mathrm{cl}}=1/2$,
\begin{eqnarray}
\epsilon_{\mathrm{ff}} & = & 0.01 e^{-\alpha_{\mathrm{vir}}} \\
\alpha_{\mathrm{vir}} & = & \frac{P/H_g}{(\pi/2) G \Sigma^2} \\
t_{\mathrm{ff}} & = & \sqrt{\frac{3\pi}{32G\rho}} \\
\rho & = & \frac{\Sigma}{2 H_g}.
\end{eqnarray}
Physically, these expressions amount to saying that half the mass at any given radius is in clouds as opposed to in a diffuse medium, roughly consistent with observations of circumnuclear starbursts \citep[e.g.,][]{rosolowsky05a}. Within those clouds, star formation proceeds at the observed rate of $\sim 1\%$ of the mass per free-fall time in virialized gas, again consistent with observations \citep{krumholz07e, krumholz12a, federrath13c, salim15a}, but that the star formation rate drops off exponentially in super-virial gas (that with $\alpha_{\mathrm{vir}} > 1$). Our definition of the virial ratio is such that $\alpha_{\mathrm{vir}}\rightarrow 1$ when gas self-gravity dominates over stellar gravity in \autoref{eq:scaleheight}, indicating that the gas is self-gravitating.

When we re-run the fiducial case with this added term, the behaviour is qualitatively unchanged, except that rather than gas accumulating indefinitely in the unstable region, the ring instead reaches a steady state where rate of star formation in the ring balances the rate of gas transport into it. This behaviour is quite insensitive to the exact star formation recipe we adopt. In reality, such a steady state is unlikely to be stable. The star formation rate surface density in the steady state is so high that the radiation flux reaches $\sim 10\%$ of the Eddington luminosity, at which point radiation pressure alone should drive rapid mass loss, even without the aid of supernovae \citep{thompson14a}. To construct our toy model, we therefore add an artificial truncation to the gas accumulation and star formation. After 22.5 Myr of evolution (chosen to be 5 Myr after the `current' CMZ as discussed in \autoref{sec:applycmz2}), when the star formation rate has reached $\sim 0.1$ ${\rm M}_\odot$ yr$^{-1}$, we simply reduce the star formation rate linearly to zero, and the gas surface density linearly to 10\% of its previous value, over a time of 4 Myr, roughly the lifetime of a massive star, in every computational cell for which $Q < 3$.\footnote{To avoid discreteness noise, we use the same trick mentioned above, whereby we interpolate $Q$ and all other quantities onto a higher resolution grid and apply this procedure to the interpolated data.} We note that this timescale is selected to give a qualitative picture, and is not quantitatively calculated.

\begin{figure}
\includegraphics[width=\hsize]{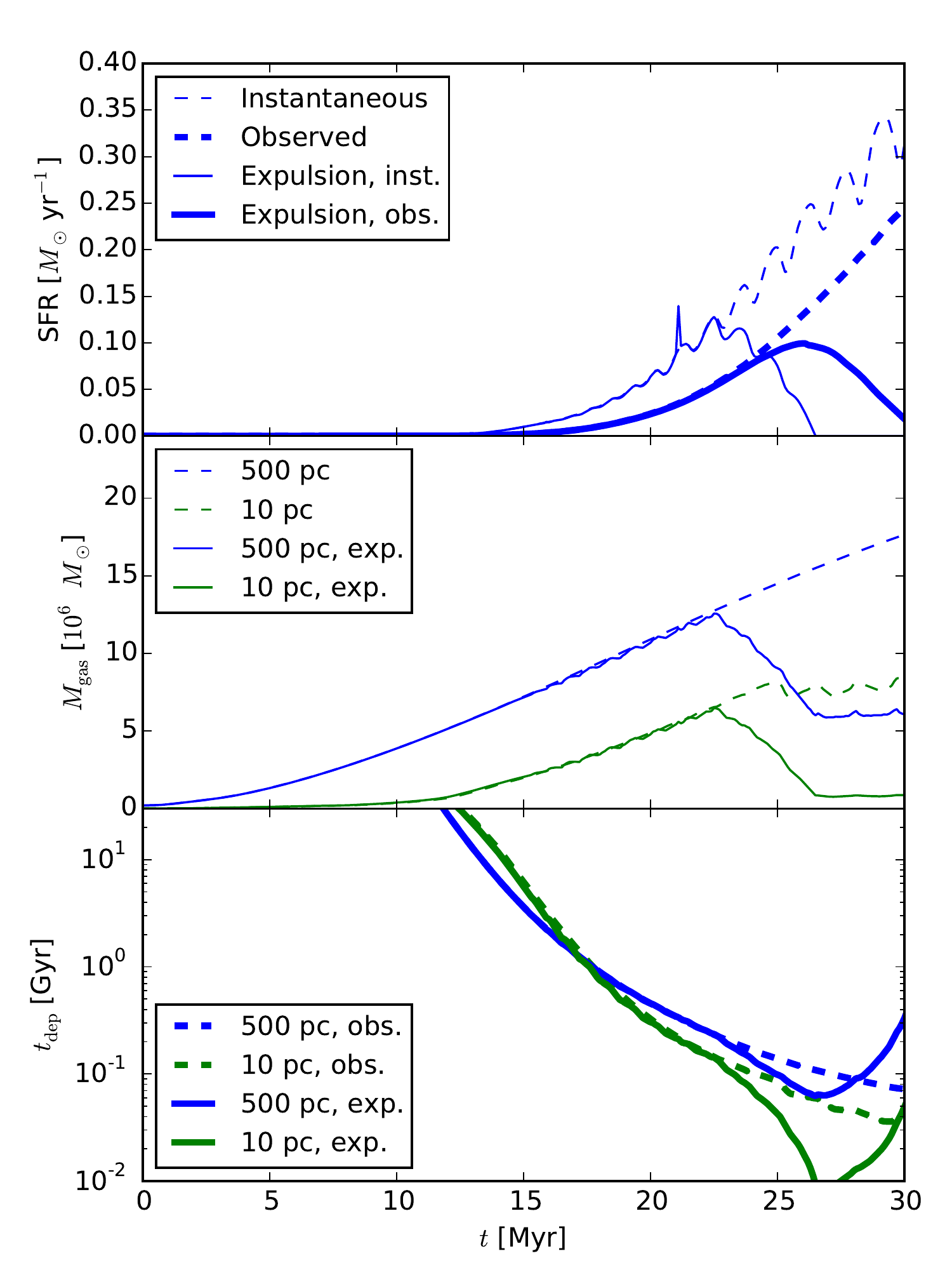}
\caption{
\label{fig:fiducial_sf}
Star formation rate (\textit{top panel}), total gas mass (\textit{middle panel}), and gas depletion time ($t_{\mathrm{dep}} = M_{\mathrm{gas}} / \mathrm{SFR}$, \textit{bottom panel}) for our fiducial run with star formation added. In all panels, quantities indicated by thin lines show true, instantaneous values, while those indicated by thick lines show values that would be inferred from observations using an ionization-sensitive star formation rate indicator such as H$\alpha$. Dashed lines show a run with no gas removal, while solid lines show the results where we use our crude gas expulsion model, whereby the star formation rate declines linearly to zero over a 4 Myr time period starting at 22.5 Myr (see main text for details). In the middle and bottom panels, the blue lines indicate values over the inner 250 pc, as would be observed with $\sim 0.5~\mathrm{kpc}$ resolution, while green lines indicate values measured only in the ring between 80 and 90 pc from the Galactic Centre, as could be observed with $\sim 10$ pc resolution. We do not include the green lines for 10 pc-resolution in the upper panel because they are essentially identical to the blue lines, since nearly all of the star formation takes place within the ring.
}
\end{figure}

With this quite crude cartoon model of feedback and gas expulsion, the overall evolution of the star formation rate, gas mass, and depletion time is as shown in \autoref{fig:fiducial_sf}. In this plot we show values computed both using all the material in the inner 250 pc, and values computed only for the material within the ring from $80 - 90$ pc ring where the gas surface density and star formation peak. Our motivation for separating these two cases is that the former is roughly analogous to what would be observed in a survey with $\sim 0.5~\mathrm{kpc}$ resolution (e.g., HERACLES, \citealt{leroy13a}), targeting a galaxy $>1~\mpc$ away using PdBI or a similar instrument; in our own Galaxy, it is roughly equivalent to considering all the material with Galactic longitude $|l| < 3^\circ$. The latter is roughly analogous to considering the material within $|l|<1^\circ$ in the Milky Way, or to the resolution that is possible in external galaxies using ALMA. We also construct a realistic ``observed" star formation rate by convolving the true star formation rate with the time-dependent ionizing luminosity $Q(\mathrm{H}^0, t)$ of a stellar population that fully samples the IMF. This correction is important because galactic centre star formation rates are usually measured with H$\alpha$ or similar ionization-based star formation rate indicators, and these effectively average the star formation rate over a time that is non-negligible compared to timescales in the model. We compute $Q(\mathrm{H}^0, t)$ using the \texttt{slug} code \citep{da-silva12a, krumholz15b} using Geneva non-rotating evolutionary tracks at Solar metallicity \citep{ekstrom12a}.

What does this evolutionary cycle look like when plotted on the usual star formation relations, which relate the star formation rate per unit area to the gas surface density, or surface density normalized by orbital time? To answer this question, we again consider observations with both $\sim0.5~\mathrm{kpc}$ and $\sim 10$ pc resolution. For the former, we take the area within a radius of 0.25~kpc (i.e.~0.2 kpc$^2$) and consider all the material within this radius \citep[cf.][]{kruijssen14b}, while for the latter we use the area of an annulus from $80-90$ pc, which is where essentially all the star formation is located -- this corresponds to roughly the central $1^\circ$ in Galactic longitude. For the orbital time, we use the orbital time for our fiducial rotation curve evaluated at the outer edge of the disc for the $\sim0.5~\mathrm{kpc}$-resolution case, and the orbital time evaluated at 90 pc for the $\sim 10$ pc-resolution case.

We must also choose when to start and stop the clock in order to make this comparison. As noted above, our initial condition contains an unrealistically small gas mass, since even a strong starburst in the unstable region seems unlikely to expel material that is $\sim 400$ pc from a galactic centre. With this consideration in mind, we choose our zero of time to be 14 Myr after the start of the fiducial run, which marks the moment when a region with $Q=1$ first appears. With this choice, and using our crude model to represent star formation feedback, gas accumulates for 8.5 Myr with a slowly increasing star formation rate. After that point gas expulsion begins and star formation slows down. The expulsion process lasts until 12.5 Myr, and thereafter the stellar population fades and the observed star formation rate (which lags the true one) declines, until gas begins re-accumulating and the cycle repeats. The full cycle in this cartoon model requires 15--20 Myr.

\begin{figure*}
\includegraphics[width=\hsize]{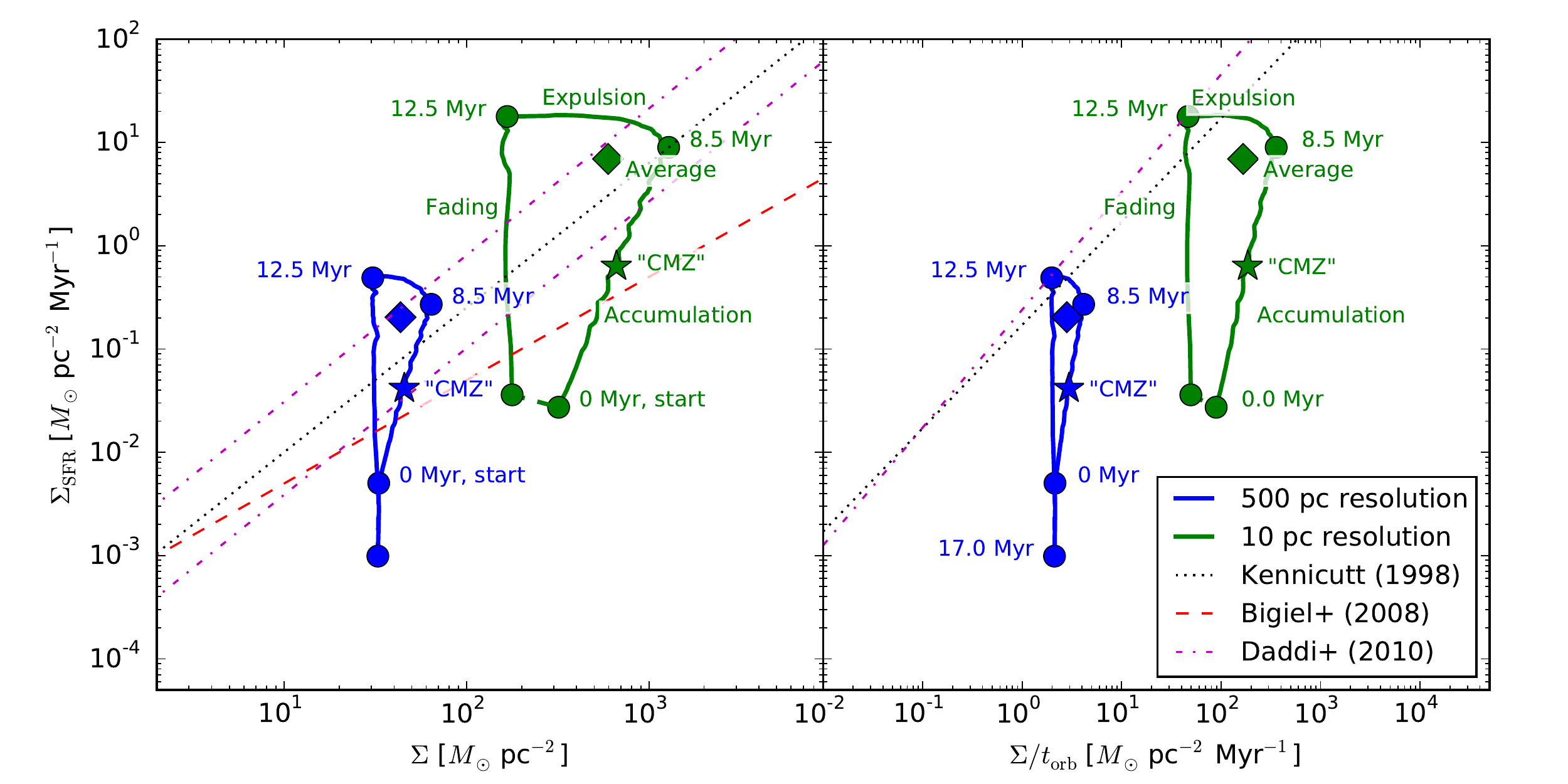}
\caption{
\label{fig:fiducial_ks}
Evolution of a CMZ in the planes of star formation surface density $\Sigma_{\mathrm{SFR}}$ versus gas surface density $\Sigma$ (\textit{left panel}) and $\Sigma_{\mathrm{SFR}}$ versus gas surface density divided by orbital time $\Sigma/t_{\mathrm{orb}}$. \textit{Blue lines} show the system as it would be observed with $\sim 0.5$ kpc resolution, while \textit{green lines} show the system at $\sim 10$ pc resolution (see text for details). \textit{Circles} mark various critical points in the evolution: the start of gas accumulation at 0 Myr (corresponding to 14 Myr of evolution in the fiducial run), the onset of gas expulsion at 8.5 Myr, the end of gas expulsion at 12.5 Myr, and complete fading of the ionizing flux from the stellar population at $\sim 17$ Myr. The \textit{stars}, labelled ``CMZ", mark the rough point in this evolutionary cycle at which the Milky Way's CMZ resides, and corresponds to 3.5 Myr after time 0 in the cycle. \textit{Diamonds} indicate the time-average over the full cycle. For comparison, we also show the star formation relations of \citet[\textit{black dotted line}]{kennicutt98a}, \citet[\textit{red dashed line}]{bigiel08a}, and \citet[\textit{magenta dot-dashed line}]{daddi10a}. For \citet{kennicutt98a}, we have multiplied the star formation rate in the original fit by $1.6$ to adjust from \citeauthor{kennicutt98a}'s adopted \citet{salpeter55a} IMF to a \citet{chabrier05a} one. For \citet{daddi10a}, the two lines shown in the left panel correspond to \citeauthor{daddi10a}'s separate fit to discs (\textit{lower line}) and starbursts (\textit{upper line}).
}
\end{figure*}

\autoref{fig:fiducial_ks} shows the cycle as it appears in the star formation relations. Clearly at either spatial resolution the variation of the CMZ's gas depletion time is dominated by the fluctuating star formation rate. The gas mass within the entire CMZ is constant to within a factor of $2-3$. The mass inside the unstable ring varies more widely, but the amount of variation is not well-determined, since it depends strongly on our assumptions about how much mass is expelled when the ring begins a starburst. Despite this uncertainty, it seems likely that the gas depletion time of the gravitationally unstable, inner CMZ will vary by at least two orders of magnitude, while the depletion time of the CMZ as a whole will show $\sim1$~dex variations. This difference arises because on large scales, the independent, outer gas reservoir is included in the depletion time measurement, which stabilises the small-scale fluctuations \citep{kruijssen14c}.

Focussing on the left panel of \autoref{fig:fiducial_ks}, and comparing the behaviour in our toy model to the star formation relations observed for entire galaxies (various lines in \autoref{fig:fiducial_ks}), we see that a CMZ measured on $\sim$kpc scales spends roughly half its time with a star formation rate below what would be expected based on its gas surface density, and about half the time at a higher star formation rate. The inner CMZ (i.e., the 100-pc stream) might seem to spend more time above the mean star formation relation, but this is somewhat misleading. The \citet{bigiel08a} relation, marked by the red dashed line in \autoref{fig:fiducial_ks}, represents a fit to disc galaxies with $\Sigma<10^2$ $\msun$ pc$^{-2}$, and does not describe the elevated star formation rates of galaxies with higher surface densities. Compared to the star formation relations observed at these higher surface densities, a CMZ observed at 10 pc-resolution also spends roughly half its time below the mean star formation rate, and half its time above. For $\sim0.5~\mathrm{kpc}$-resolution observations, the scatter about the conventional star formation relations is $\sim 1-1.5$ dex, while at $\sim 10$ pc resolution it is as much as $\sim 2-2.5$ dex, though the exact value depends on the details of gas expulsion. The present-day CMZ of the Milky Way is in the phase of its evolution when it lies below the mean star formation relations.

The above comparison changes when dividing the gas surface density by the orbital time (right-hand panel in \autoref{fig:fiducial_ks}). In that case, the extremely short orbital time of a CMZ places it below all star formation relations {\it except} during its starbursts. This holds for both the inner CMZ and the entire CMZ. The division by the orbital time therefore provides an efficient way of highlighting galactic centres at a star formation minimum, irrespective of the spatial resolution.

Finally, we note that in all four discussed cases (two spatial scales in two panels of \autoref{fig:fiducial_ks}), the cycle-averaged CMZ agrees with \citet{daddi10a}'s observed star formation relations for starburst galaxies to within a factor of 3. Without any significant evolution of the rotation curve or the gas inflow rate (either by secular processes or due to external perturbations), the CMZ should continue to evolve through the cycle shown in \autoref{fig:fiducial_ks}. If the conditions do change fundamentally, then this will affect the time-scales of the different phases. The cycle is not expected to shut off altogether, because there should always be a radius interval where the rotation curve has a near-solid body part and gas accumulates.

\subsection{Predictions for the Centres of External Galaxies}
The model presented here provides the first self-consistent theory that quantitatively reproduces the main features of the CMZ of the Milky Way. It is therefore desirable to make predictions that hold for galactic centres of barred spiral galaxies in general.

In summary, our model consists of the following critical ingredients.
\begin{enumerate}
\item\label{pt:ilr}
The galaxy has an ILR to which gas is supplied by the bar at sufficiently high velocity dispersions ($\sigma_{\rm 1D}>10$--$15~\kms$) to drive angular momentum transport and thus prevent the gas inflow from stalling.
\item\label{pt:rotcurve}
For most of the radial range within the ILR, the rotation curve is near-flat so that the shear is high ($1-\beta\ga0.8$).
\item\label{pt:ai}
As a result of points~\ref{pt:ilr} and~\ref{pt:rotcurve}, acoustic instabilities develop in the gas after it enters the ILR. These instabilities drive efficient angular momentum transport, leading to inflowing gas with high turbulent pressures ($P/k>10^5~{\rm K}~\cmc$), extreme gravitational stability ($Q>10$), and a low star formation rate ($t_{\rm dep}\sim10~\gyr$).
\item\label{pt:shear}
Within the ILR, there is a radius where the rotation curve transitions from near-flat to near-solid body, causing a minimum in the shear ($1-\beta\la0.4$). The angular momentum transport and mass inflow both stall at this radius.
\item\label{pt:density}
At this radius, the inwards mass flux is sufficient for the accumulating gas to eventually exceed the stellar density ($\rho>\rho_\star$).
\item\label{pt:gi}
As a result of points~\ref{pt:shear} and~\ref{pt:density}, gravitational instabilities develop in the growing gas reservoir near the shear minimum. At the high gas densities required by point~\ref{pt:density}, these instabilities drive the gas into rapid free-fall ($t_{\rm ff}\la1~\myr$), resulting in a starburst ($t_{\rm dep}\la0.1~\gyr$).
\end{enumerate}
These conditions are commonly satisfied in the centres of barred spiral galaxies \citep[e.g.][]{elmegreen02b,jogee02a,martini03a,jogee05a,leroy08a,sandstrom10a,nesvadba11a,sani12a}. As a result, the physical processes described in this paper and the qualitative cycle of \autoref{fig:fiducial_ks} (also see Figure~6 of \citealt{kruijssen14b}) should be common in other galactic centres.

While our model may apply qualitatively to other galaxies, it is unclear if the quantitative predictions hold. Given the close relation between the Galactic rotation curve and the predictions of our model, it is to be expected that the details of extragalactic rotation curves lead to quantitatively different radial profiles of, e.g., the gas surface density, velocity dispersion, and mass inflow rate. However, the duty cycle from quiescence to starburst activity and back may be relatively unaffected. The ILR of more massive galaxies may reside at larger radii (which at first sight implies longer gas accretion time-scales), but the deeper gravitational potential and the correspondingly increased gas pressures will accelerate the angular momentum transport by acoustic instabilities (which decreases the gas accretion time-scale again). As a result, it is possible that the duty cycle predicted for the CMZ (see \autoref{fig:fiducial_ks}) has a broader application than perhaps naively expected. From an empirical perspective, it is therefore interesting to make a simple comparison to the population statistics of other galactic centres.

A key result of the discussion in \autoref{sec:applycmz3} is that the details of the star formation cycle and its position relative to empirical star formation relations depend on the spatial resolution (compare the blue and green tetragons in \autoref{fig:fiducial_ks}). Low-resolution observations should find a smaller variation of the gas depletion time than high-resolution observations. \citet[Figure~13]{leroy13a} find a $\sim1$~dex scatter of the gas depletion time in the central $R<0.5~\kpc$ of 30 nearby disc galaxies from the HERACLES survey. A similar range is found for other galaxy samples \citep[e.g.,][]{sakamoto99a,jogee05a,hsieh11a,sani12a,saintonge12a,fisher13a}. In addition, the galactic centres in the HERACLES sample exhibit a $\sim0.3$~dex decrease of the gas depletion time relative to galactic discs \citep{leroy13a}. Both the observed scatter and deviation of the depletion time present a remarkably good match to our predictions for the same area ($|l|<3^\circ$ and $0.5~\kpc$ resolution).

In view of the good agreement between our model and the galactic centres in the HERACLES sample, we can make a prediction for the relative frequency of quiescent, normal and starbursting galactic centres across the population of barred spiral galaxies. Of course, this assumes that our CMZ model describes the duty cycle despite obvious differences in rotation curves and other galaxy properties. In turn, large observational samples of galactic centres can be used to test this assumption.

We estimate the relative occurrence rates by using the time-scales associated to each of the phases in the evolutionary cycle from \autoref{sec:applycmz3}. For $0.5~\kpc$-resolution observations, we predict that $\sim 1/3$ of all galaxies have a `normal' gas depletion time-scale of $\sim 2$ Gyr (the \citet{bigiel08a} line in \autoref{fig:fiducial_ks}) within a factor of 2, whereas $\sim 2/3$ have elevated star formation activity and shorter gas depletion times; a small fraction have depressed star formation rates. This is again consistent with the distribution of gas depletion times found by \citet{leroy13a}. By contrast, for high-resolution ($\sim10~\pc$) observations, we predict that the depletion times will almost always be shorter than the $\sim 2$ Gyr found at larger galactic radii, and that $\sim 1/2$ of the time they will also lie above the star formation relations of \citet{kennicutt98a} or \citet{daddi10a} for ``normal" (i.e., non-starbursting) galaxies. This is a fundamentally different distribution than at low resolution, and shows that upcoming ALMA observations provide a unique opportunity to test our predictions.

We reiterate one of the conclusions from \autoref{sec:applycmz3}, that dividing the gas surface density by the orbital time (which is naturally short in galactic centres) shifts galactic centres below the relation followed by galaxy discs at any spatial resolution at almost all times (see the right-hand panel of \autoref{fig:fiducial_ks}). Only in outburst do these star formation relations approach the usual ones found at larger galactic radii. This provides a way of highlighting galactic centres with low star formation rates.

More direct tests of our model will be enabled by high-resolution ALMA observations of nearby galactic centres. Such observations provide the necessary input for our model (i.e., high-resolution rotation curves) as well as the predicted observables (e.g., radial profiles of the gas surface density, velocity dispersion, mass inflow rate, dense gas fraction, etc.). Next to building up a statistically representative sample of galactic centres, such observations have the additional advantage that the centres of other barred spiral galaxies may not be at the same phase of the evolutionary cycle as the Galactic CMZ. That way, they provide the best opportunity to characterise the relevant physics during all evolutionary phases and to derive an accurate duty cycle for star formation activity in the centres of barred spiral galaxies.

\subsection{\red{Relation to Previous and Future Theoretical Work}}

\red{We end this discussion by commenting on how the model we present here relates to other theoretical work on the dynamics of gas in galactic CMZs. There have been a number of previous hydrodynamic simulations of such regions \citep[e.g.,][]{rodriguez-fernandez08a, baba10a, pettitt14a, cole14a, sormani15a}. Most of this work has been focused either on explaining the observed kinematics of gas near the Galactic center, or on studying how a bar can drive mass inward and create circumnuclear structures. The picture that these simulations present is that a bar, by creating self-intersecting $x_1$ orbits, leads the gas to shock and plunge inward, whereupon it finds itself on $x_2$ orbits at distances of a few hundred pc from the Galactic center. This behavior is roughly consistent with the predictions of analytic models \citep{binney91a}.}

\red{
With the exception of \citet{sormani15a}, who reach $5$ pc resolution, none of the previous works of which we are aware have enough resolution to model the $\sim 100$ pc size-scale structures on which we focus here. Because they are focused on larger scales, then tend to have resolutions of $\sim 50$ pc at best.\footnote{\red{\citet{rodriguez-fernandez08a} use a sticky-particle method with a $\sim 5$ pc collision radius, but the point still holds -- because they are not actually simulating hydrodynamics, their method is not capable of representing an accretion disc.}} \citeauthor{sormani15a} find that the gas on $x_1$ orbits that shocks does subsequently settle into a disc-like structure a few hundred pc in size, with the exact settling radius in their simulations depending on their assumed gas equation of state. Given the relatively small wavelength of the acoustic instability (cf.~\autoref{fig:wavelength}), they unfortunately cannot resolve it, but in principle similar simulations at higher resolution should be able to model the scenario that we propose here. 
}

\red{
More broadly, future simulations would provide a valuable test of the central assumption in our picture. We have assumed that, once gas is transferred inward to $x_2$ orbits by shocks, we can approximate its behavior as an axisymmetric disc subject to perturbations from the bar. This approach emphasizes the gaseous nature of the flow in the inner parts of a galaxy. In contrast, many previous authors have instead chosen to visualize the gas in this region as consisting of effectively-collisionless clouds moving on ballistic orbits \citep[e.g.,][]{binney91a, sofue95a}, which plunge into the central black hole on an orbital timescale. If this latter picture is closer to the truth, then our disc-based picture is called into question. While \citet{sormani15a}'s simulation certainly does suggest the formation of a disc-like structure, the question of how well gas inside the $x_1$ region can be approximated as a perturbed disc remains unsettled.
}

\section{Conclusion}
\label{sec:conclusion}

In this paper we present the first global model for gas transport and star formation in the centres of barred spiral galaxies, focusing on the Milky Way's Central Molecular Zone (CMZ) as a paradigmatic example. The model is based on a few simple ingredients. First, the rotation curves of galaxies, which are generally close to flat at large radii, must eventually turn over and approach solid body near galactic centres, producing a minimum in the local shear. In the CMZ, this minimum occurs $\sim 100$ pc from the Galactic centre. Second, gas that orbits in the flat rotation curve part of the potential will be unstable due to acoustic instabilities pumped within the bar's inner Linbdlad resonance (ILR). These instabilities will transport angular momentum and cause mass to flow inward, until the flow stagnates in the low-shear region where angular momentum transport is suppressed. In our axisymmetric models this produces a ring-like structure, though in a real galaxy the ring is likely to be partially rather than completely filled. Third, within the region of rapid infow, the velocity dispersion driven by the instability render the gas highly-supervirial, and thus unable to undergo collapse and star formation. In contrast, the accumulating gas in the stagnation region must eventually become self-gravitating and undergo vigorous star formation, likely leading to a blowout that removes much of the gas. The cycle will then begin again as new mass accumulates.

The scenario we propose, which is supported by numerical calculations of disc evolution performed with the \texttt{VADER} code \citep{krumholz15a}, naturally explains a wide range of observations for both the Milky Way CMZ and other galactic centres. In the Milky Way, our model naturally produces the observed ring-like stream of gas at $\sim 100$ pc from the Galactic centre. This prediction requires no fine-tuning, and follows simply from the observed shape of the Galactic rotation curve. Second, our model naturally explains the broad diversity of star formation rates seen in the centres of neraby galaxies, some of which have gas depletion times comparable to those found at larger galactocentric radii, and some of which show significantly faster star formation. In our model, such diversity occurs naturally as a result of the cycle of accumulation, starburst, gas clearing, and fading that rapid transport produces. It thereby quantifies and greatly expands the scenario of episodic star formation proposed by \citet{kruijssen14b}.

In addition to explaining existing observations, this model makes specific predictions for both the morphology and the quantitative star formation behaviour of the centres of other galaxies when observed at high resolution. While the requisite observations have not yet been made, they are well within the reach of ALMA. A program to investigate the star formation behaviour of nearby galactic centres at high resolution should therefore be a high priority in the coming ALMA cycles. Such observations would allow us to observe galactic centres that are at very different phases of the boom and bust cycle than our own relatively quiescent one. A sufficiently large sample would even allow us to place statistical constraints on the duration of the cycle. This in turn would provide strong confirmation that, at least in the most extreme environments, star formation is more than a purely local process that proceeds independent of its galactic environment.

\section*{Acknowledgements}

We thank Steve Longmore for helpful discussions and suggestions. This work was supported by NSF grants AST-0955300 and AST-1405962 to MRK. The authors thank the Aspen Center for Physics, which is supported by NSF Grant PHY-1066293, for its hospitality during the early phases of this work.

\bibliographystyle{mn2e}
\bibliography{refs}

\begin{thebibliography}{138}
\expandafter\ifx\csname natexlab\endcsname\relax\def\natexlab#1{#1}\fi

\bibitem[{Akima(1970)}]{akima70a}
Akima H., 1970, \jacm, 17, 589

\bibitem[{{Andr{\'e}} {et~al}\mbox{.}(2007){Andr{\'e}}, {Belloche}, {Motte}, \&
  {Peretto}}]{andre07a}
{Andr{\'e}} P., {Belloche} A., {Motte} F., {Peretto} N., 2007, \aap, 472, 519

\bibitem[{{Antoja} {et~al}\mbox{.}(2014){Antoja}, {Helmi}, {Dehnen},
  {Bienaym{\'e}}, {Bland-Hawthorn}, {Famaey}, {Freeman}, {Gibson}, {Gilmore},
  {Grebel}, {Kordopatis}, {Kunder}, {Minchev}, {Munari}, {Navarro}, {Parker},
  {Reid}, {Seabroke}, {Siebert}, {Steinmetz}, {Watson}, {Wyse}, \&
  {Zwitter}}]{antoja14a}
{Antoja} T. {et~al.}, 2014, \aap, 563, A60

\bibitem[{{Ao} {et~al}\mbox{.}(2013){Ao}, {Henkel}, {Menten}, {Requena-Torres},
  {Stanke}, {Mauersberger}, {Aalto}, {M{\"u}hle}, \& {Mangum}}]{ao13a}
{Ao} Y. {et~al.}, 2013, \aap, 550, A135

\bibitem[{{Baba}, {Saitoh} \& {Wada}(2010){Baba}, {Saitoh}, \&
  {Wada}}]{baba10a}
{Baba} J., {Saitoh} T.~R., {Wada} K., 2010, \pasj, 62, 1413

\bibitem[{{Bally} {et~al}\mbox{.}(2010){Bally}, {Aguirre}, {Battersby},
  {Bradley}, {Cyganowski}, {Dowell}, {Drosback}, {Dunham}, {Evans}, {Ginsburg},
  {Glenn}, {Harvey}, {Mills}, {Merello}, {Rosolowsky}, {Schlingman}, {Shirley},
  {Stringfellow}, {Walawender}, \& {Williams}}]{bally10a}
{Bally} J. {et~al.}, 2010, \apj, 721, 137

\bibitem[{{Bally} {et~al}\mbox{.}(2014){Bally}, {Rathborne}, {Longmore},
  {Jackson}, {Alves}, {Bressert}, {Contreras}, {Foster}, {Garay}, {Ginsburg},
  {Johnston}, {Kruijssen}, {Testi}, \& {Walsh}}]{bally14a}
{Bally} J. {et~al.}, 2014, \apj, 795, 28

\bibitem[{{Bally} {et~al}\mbox{.}(1988){Bally}, {Stark}, {Wilson}, \&
  {Henkel}}]{bally88a}
{Bally} J., {Stark} A.~A., {Wilson} R.~W., {Henkel} C., 1988, \apj, 324, 223

\bibitem[{{Bertin} {et~al}\mbox{.}(1989){Bertin}, {Lin}, {Lowe}, \&
  {Thurstans}}]{bertin89a}
{Bertin} G., {Lin} C.~C., {Lowe} S.~A., {Thurstans} R.~P., 1989, \apj, 338, 104

\bibitem[{{Bhattacharjee}, {Chaudhury} \& {Kundu}(2014){Bhattacharjee},
  {Chaudhury}, \& {Kundu}}]{bhattacharjee14a}
{Bhattacharjee} P., {Chaudhury} S., {Kundu} S., 2014, \apj, 785, 63

\bibitem[{{Bigiel} {et~al}\mbox{.}(2008){Bigiel}, {Leroy}, {Walter}, {Brinks},
  {de Blok}, {Madore}, \& {Thornley}}]{bigiel08a}
{Bigiel} F., {Leroy} A., {Walter} F., {Brinks} E., {de Blok} W.~J.~G., {Madore}
  B., {Thornley} M.~D., 2008, \aj, 136, 2846

\bibitem[{{Binney} {et~al}\mbox{.}(1991){Binney}, {Gerhard}, {Stark}, {Bally},
  \& {Uchida}}]{binney91a}
{Binney} J., {Gerhard} O.~E., {Stark} A.~A., {Bally} J., {Uchida} K.~I., 1991,
  \mnras, 252, 210

\bibitem[{{Binney} \& {Tremaine}(1987)}]{binney87a}
{Binney} J., {Tremaine} S., 1987, Galactic Dynamics. Princeton University
  Press, Princeton, NJ

\bibitem[{{Bland-Hawthorn} \& {Cohen}(2003)}]{bland-hawthorn03a}
{Bland-Hawthorn} J., {Cohen} M., 2003, \apj, 582, 246

\bibitem[{{Bournaud}, {Elmegreen} \& {Elmegreen}(2007){Bournaud}, {Elmegreen},
  \& {Elmegreen}}]{bournaud07a}
{Bournaud} F., {Elmegreen} B.~G., {Elmegreen} D.~M., 2007, \apj, 670, 237

\bibitem[{{Cacciato}, {Dekel} \& {Genel}(2012){Cacciato}, {Dekel}, \&
  {Genel}}]{cacciato12a}
{Cacciato} M., {Dekel} A., {Genel} S., 2012, \mnras, 421, 818

\bibitem[{{Carretti} {et~al}\mbox{.}(2013){Carretti}, {Crocker},
  {Staveley-Smith}, {Haverkorn}, {Purcell}, {Gaensler}, {Bernardi}, {Kesteven},
  \& {Poppi}}]{carretti13a}
{Carretti} E. {et~al.}, 2013, \nat, 493, 66

\bibitem[{{Ceverino} {et~al}\mbox{.}(2015){Ceverino}, {Dekel}, {Tweed}, \&
  {Primack}}]{ceverino15a}
{Ceverino} D., {Dekel} A., {Tweed} D., {Primack} J., 2015, \mnras, 447, 3291

\bibitem[{{Chabrier}(2005)}]{chabrier05a}
{Chabrier} G., 2005, in Astrophysics and Space Science Library, Vol. 327, The
  Initial Mass Function 50 Years Later, {Corbelli} E., {Palla} F., {Zinnecker}
  H., eds., Springer, Dordrecht, pp. 41--+

\bibitem[{{Clark} {et~al}\mbox{.}(2013){Clark}, {Glover}, {Ragan}, {Shetty}, \&
  {Klessen}}]{clark13b}
{Clark} P.~C., {Glover} S.~C.~O., {Ragan} S.~E., {Shetty} R., {Klessen} R.~S.,
  2013, \apjl, 768, L34

\bibitem[{{Cole} {et~al}\mbox{.}(2014){Cole}, {Debattista}, {Erwin}, {Earp}, \&
  {Ro{\v s}kar}}]{cole14a}
{Cole} D.~R., {Debattista} V.~P., {Erwin} P., {Earp} S.~W.~F., {Ro{\v s}kar}
  R., 2014, \mnras, 445, 3352

\bibitem[{{Colombo} {et~al}\mbox{.}(2014){Colombo}, {Hughes}, {Schinnerer},
  {Meidt}, {Leroy}, {Pety}, {Dobbs}, {Garc{\'{\i}}a-Burillo}, {Dumas},
  {Thompson}, {Schuster}, \& {Kramer}}]{colombo14a}
{Colombo} D. {et~al.}, 2014, \apj, 784, 3

\bibitem[{{Crocker}(2012)}]{crocker12a}
{Crocker} R.~M., 2012, \mnras, 423, 3512

\bibitem[{{Daddi} {et~al}\mbox{.}(2010){Daddi}, {Elbaz}, {Walter}, {Bournaud},
  {Salmi}, {Carilli}, {Dannerbauer}, {Dickinson}, {Monaco}, \&
  {Riechers}}]{daddi10a}
{Daddi} E. {et~al.}, 2010, \apjl, 714, L118

\bibitem[{{Debattista}, {Gerhard} \& {Sevenster}(2002){Debattista}, {Gerhard},
  \& {Sevenster}}]{debattista02a}
{Debattista} V.~P., {Gerhard} O., {Sevenster} M.~N., 2002, \mnras, 334, 355

\bibitem[{Dimonte {et~al}\mbox{.}(2004)Dimonte, Youngs, Dimits, Weber, Marinak,
  Wunsch, Garasi, Robinson, Andrews, Ramaprabhu, Calder, Fryxell, Biello,
  Dursi, MacNeice, Olson, Ricker, Rosner, Timmes, Tufo, Young, \&
  Zingale}]{dimonte04a}
Dimonte G. {et~al.}, 2004, Physics of Fluids, 16, 1668

\bibitem[{{Dobbs} {et~al}\mbox{.}(2014){Dobbs}, {Krumholz},
  {Ballesteros-Paredes}, {Bolatto}, {Fukui}, {Heyer}, {Low}, {Ostriker}, \&
  {V{\'a}zquez-Semadeni}}]{dobbs14a}
{Dobbs} C.~L. {et~al.}, 2014, Protostars and Planets VI, 3

\bibitem[{{Ekstr{\"o}m} {et~al}\mbox{.}(2012){Ekstr{\"o}m}, {Georgy},
  {Eggenberger}, {Meynet}, {Mowlavi}, {Wyttenbach}, {Granada}, {Decressin},
  {Hirschi}, {Frischknecht}, {Charbonnel}, \& {Maeder}}]{ekstrom12a}
{Ekstr{\"o}m} S. {et~al.}, 2012, \aap, 537, A146

\bibitem[{{Elmegreen}, {Elmegreen} \& {Eberwein}(2002){Elmegreen}, {Elmegreen},
  \& {Eberwein}}]{elmegreen02b}
{Elmegreen} D.~M., {Elmegreen} B.~G., {Eberwein} K.~S., 2002, \apj, 564, 234

\bibitem[{{Faucher-Gigu{\`e}re}, {Quataert} \&
  {Hopkins}(2013){Faucher-Gigu{\`e}re}, {Quataert}, \&
  {Hopkins}}]{faucher-giguere13a}
{Faucher-Gigu{\`e}re} C.-A., {Quataert} E., {Hopkins} P.~F., 2013, \mnras, 433,
  1970

\bibitem[{{Federrath}(2013)}]{federrath13c}
{Federrath} C., 2013, \mnras, 436, 3167

\bibitem[{{Federrath} \& {Banerjee}(2015)}]{federrath15a}
{Federrath} C., {Banerjee} S., 2015, \mnras, 448, 3297

\bibitem[{{Federrath} \& {Klessen}(2012)}]{federrath12a}
{Federrath} C., {Klessen} R.~S., 2012, \apj, 761, 156

\bibitem[{{Federrath} \& {Klessen}(2013)}]{federrath13a}
{Federrath} C., {Klessen} R.~S., 2013, \apj, 763, 51

\bibitem[{{Federrath}, {Klessen} \& {Schmidt}(2008){Federrath}, {Klessen}, \&
  {Schmidt}}]{federrath08a}
{Federrath} C., {Klessen} R.~S., {Schmidt} W., 2008, \apjl, 688, L79

\bibitem[{{Fisher} {et~al}\mbox{.}(2013){Fisher}, {Bolatto}, {Drory}, {Combes},
  {Blitz}, \& {Wong}}]{fisher13a}
{Fisher} D.~B., {Bolatto} A., {Drory} N., {Combes} F., {Blitz} L., {Wong} T.,
  2013, \apj, 764, 174

\bibitem[{{Forbes}, {Krumholz} \& {Burkert}(2012){Forbes}, {Krumholz}, \&
  {Burkert}}]{forbes12a}
{Forbes} J., {Krumholz} M., {Burkert} A., 2012, \apj, 754, 48

\bibitem[{{Forbes} {et~al}\mbox{.}(2014){Forbes}, {Krumholz}, {Burkert}, \&
  {Dekel}}]{forbes14a}
{Forbes} J.~C., {Krumholz} M.~R., {Burkert} A., {Dekel} A., 2014, \mnras, 438,
  1552

\bibitem[{{Galassi} {et~al}\mbox{.}(2009){Galassi}, {Davies}, {Theiler},
  {Gough}, {Jungman}, {Alken}, {Booth}, \& {Rossi}}]{galassi09a}
{Galassi} M., {Davies} J., {Theiler} J., {Gough} B., {Jungman} G., {Alken} P.,
  {Booth} M., {Rossi} F., 2009, GNU Scientific Library Reference Manual, 3rd
  edn. ISBN 0954612078

\bibitem[{{Gans} \& {Gill}(1984)}]{gans84a}
{Gans} P., {Gill} J.~B., 1984, Applied Spectroscopy, 38, 370

\bibitem[{{Ginsburg} {et~al}\mbox{.}(2015){Ginsburg}, {Henkel}, {Ao},
  {Riquelme}, {Kauffmann}, {Pillai}, {Mills}, {Requena-Torres}, {Immer},
  {Testi}, {Ott}, {Bally}, {Battersby}, {Darling}, {Aalto}, {Stanke},
  {Kendrew}, {Kruijssen}, {Longmore}, {Dale}, {Guesten}, \&
  {Menten}}]{ginsburg15a}
{Ginsburg} A. {et~al.}, 2015, \aap~submitted

\bibitem[{{Goldreich} \& {Lynden-Bell}(1965)}]{goldreich65a}
{Goldreich} P., {Lynden-Bell} D., 1965, \mnras, 130, 125

\bibitem[{{Hennebelle} \& {Chabrier}(2013)}]{hennebelle13a}
{Hennebelle} P., {Chabrier} G., 2013, \apj, 770, 150

\bibitem[{{Hopkins} \& {Quataert}(2010)}]{hopkins10b}
{Hopkins} P.~F., {Quataert} E., 2010, \mnras, 407, 1529

\bibitem[{{Hopkins}, {Quataert} \& {Murray}(2011){Hopkins}, {Quataert}, \&
  {Murray}}]{hopkins11a}
{Hopkins} P.~F., {Quataert} E., {Murray} N., 2011, \mnras, 417, 950

\bibitem[{{Hsieh} {et~al}\mbox{.}(2011){Hsieh}, {Matsushita}, {Liu}, {Ho},
  {Oi}, \& {Wu}}]{hsieh11a}
{Hsieh} P.-Y., {Matsushita} S., {Liu} G., {Ho} P.~T.~P., {Oi} N., {Wu} Y.-L.,
  2011, \apj, 736, 129

\bibitem[{{Hughes} {et~al}\mbox{.}(2013){Hughes}, {Meidt}, {Colombo},
  {Schinnerer}, {Pety}, {Leroy}, {Dobbs}, {Garc{\'{\i}}a-Burillo}, {Thompson},
  {Dumas}, {Schuster}, \& {Kramer}}]{hughes13a}
{Hughes} A. {et~al.}, 2013, \apj, 779, 46

\bibitem[{{Jogee}, {Scoville} \& {Kenney}(2005){Jogee}, {Scoville}, \&
  {Kenney}}]{jogee05a}
{Jogee} S., {Scoville} N., {Kenney} J.~D.~P., 2005, \apj, 630, 837

\bibitem[{{Jogee} {et~al}\mbox{.}(2002){Jogee}, {Shlosman}, {Laine},
  {Englmaier}, {Knapen}, {Scoville}, \& {Wilson}}]{jogee02a}
{Jogee} S., {Shlosman} I., {Laine} S., {Englmaier} P., {Knapen} J.~H.,
  {Scoville} N., {Wilson} C.~D., 2002, \apj, 575, 156

\bibitem[{{Kauffmann}, {Pillai} \& {Zhang}(2013){Kauffmann}, {Pillai}, \&
  {Zhang}}]{kauffmann13a}
{Kauffmann} J., {Pillai} T., {Zhang} Q., 2013, \apjl, 765, L35

\bibitem[{{Kennicutt}(1998)}]{kennicutt98a}
{Kennicutt}, Jr. R.~C., 1998, \apj, 498, 541

\bibitem[{{Kirk}, {Johnstone} \& {Tafalla}(2007){Kirk}, {Johnstone}, \&
  {Tafalla}}]{kirk07a}
{Kirk} H., {Johnstone} D., {Tafalla} M., 2007, \apj, 668, 1042

\bibitem[{{Koepferl} {et~al}\mbox{.}(2015){Koepferl}, {Robitaille}, {Morales},
  \& {Johnston}}]{koepferl15a}
{Koepferl} C.~M., {Robitaille} T.~P., {Morales} E.~F.~E., {Johnston} K.~G.,
  2015, \apj, 799, 53

\bibitem[{{Kormendy} \& {Kennicutt}(2004)}]{kormendy04a}
{Kormendy} J., {Kennicutt}, Jr. R.~C., 2004, \araa, 42, 603

\bibitem[{{Kratter} {et~al}\mbox{.}(2010){Kratter}, {Matzner}, {Krumholz}, \&
  {Klein}}]{kratter10a}
{Kratter} K.~M., {Matzner} C.~D., {Krumholz} M.~R., {Klein} R.~I., 2010, \apj,
  708, 1585

\bibitem[{{Kruijssen}, {Dale} \& {Longmore}(2015){Kruijssen}, {Dale}, \&
  {Longmore}}]{kruijssen15a}
{Kruijssen} J.~M.~D., {Dale} J.~E., {Longmore} S.~N., 2015, \mnras, 447, 1059

\bibitem[{{Kruijssen} \& {Longmore}(2013)}]{kruijssen13a}
{Kruijssen} J.~M.~D., {Longmore} S.~N., 2013, \mnras, 435, 2598

\bibitem[{{Kruijssen} \& {Longmore}(2014)}]{kruijssen14c}
{Kruijssen} J.~M.~D., {Longmore} S.~N., 2014, \mnras, 439, 3239

\bibitem[{{Kruijssen} {et~al}\mbox{.}(2014){Kruijssen}, {Longmore},
  {Elmegreen}, {Murray}, {Bally}, {Testi}, \& {Kennicutt}}]{kruijssen14b}
{Kruijssen} J.~M.~D., {Longmore} S.~N., {Elmegreen} B.~G., {Murray} N., {Bally}
  J., {Testi} L., {Kennicutt} R.~C., 2014, \mnras, 440, 3370

\bibitem[{{Krumholz}(2013)}]{krumholz13c}
{Krumholz} M.~R., 2013, \mnras, 436, 2747

\bibitem[{{Krumholz}(2014)}]{krumholz14c}
{Krumholz} M.~R., 2014, \physrep, 539, 49

\bibitem[{{Krumholz} \& {Burkert}(2010)}]{krumholz10c}
{Krumholz} M.~R., {Burkert} A., 2010, \apj, 724, 895

\bibitem[{{Krumholz}, {Dekel} \& {McKee}(2012){Krumholz}, {Dekel}, \&
  {McKee}}]{krumholz12a}
{Krumholz} M.~R., {Dekel} A., {McKee} C.~F., 2012, \apj, 745, 69

\bibitem[{{Krumholz} \& {Forbes}(2015)}]{krumholz15a}
{Krumholz} M.~R., {Forbes} J.~C., 2015, Astronomy and Computing, 11, 1

\bibitem[{{Krumholz} {et~al}\mbox{.}(2015){Krumholz}, {Fumagalli}, {da Silva},
  {Rendahl}, \& {Parra}}]{krumholz15b}
{Krumholz} M.~R., {Fumagalli} M., {da Silva} R.~L., {Rendahl} T., {Parra} J.,
  2015, \mnras, in press, arXiv:1502.05408

\bibitem[{{Krumholz} \& {McKee}(2005)}]{krumholz05c}
{Krumholz} M.~R., {McKee} C.~F., 2005, \apj, 630, 250

\bibitem[{{Krumholz}, {McKee} \& {Tumlinson}(2009){Krumholz}, {McKee}, \&
  {Tumlinson}}]{krumholz09b}
{Krumholz} M.~R., {McKee} C.~F., {Tumlinson} J., 2009, \apj, 699, 850

\bibitem[{{Krumholz} \& {Tan}(2007)}]{krumholz07e}
{Krumholz} M.~R., {Tan} J.~C., 2007, \apj, 654, 304

\bibitem[{{Lada} {et~al}\mbox{.}(2012){Lada}, {Forbrich}, {Lombardi}, \&
  {Alves}}]{lada12a}
{Lada} C.~J., {Forbrich} J., {Lombardi} M., {Alves} J.~F., 2012, \apj, 745, 190

\bibitem[{{Lada} \& {Lada}(2003)}]{lada03a}
{Lada} C.~J., {Lada} E.~A., 2003, \araa, 41, 57

\bibitem[{{Lau} \& {Bertin}(1978)}]{lau78a}
{Lau} Y.~Y., {Bertin} G., 1978, \apj, 226, 508

\bibitem[{{Launhardt}, {Zylka} \& {Mezger}(2002){Launhardt}, {Zylka}, \&
  {Mezger}}]{launhardt02a}
{Launhardt} R., {Zylka} R., {Mezger} P.~G., 2002, \aap, 384, 112

\bibitem[{{Lemaster} \& {Stone}(2009)}]{lemaster09a}
{Lemaster} M.~N., {Stone} J.~M., 2009, \apj, 691, 1092

\bibitem[{{Leroy} {et~al}\mbox{.}(2014){Leroy}, {Bolatto}, {Ostriker},
  {Rosolowsky}, {Walter}, {Warren}, {Donovan Meyer}, {Hodge}, {Meier}, {Ott},
  {Sandstrom}, {Schruba}, {Veilleux}, \& {Zwaan}}]{leroy14a}
{Leroy} A.~K. {et~al.}, 2014, \apj, in press, arXiv:1411.2836

\bibitem[{{Leroy} {et~al}\mbox{.}(2008){Leroy}, {Walter}, {Brinks}, {Bigiel},
  {de Blok}, {Madore}, \& {Thornley}}]{leroy08a}
{Leroy} A.~K., {Walter} F., {Brinks} E., {Bigiel} F., {de Blok} W.~J.~G.,
  {Madore} B., {Thornley} M.~D., 2008, \aj, 136, 2782

\bibitem[{{Leroy} {et~al}\mbox{.}(2013){Leroy}, {Walter}, {Sandstrom},
  {Schruba}, {Munoz-Mateos}, {Bigiel}, {Bolatto}, {Brinks}, {de Blok}, {Meidt},
  {Rix}, {Rosolowsky}, {Schinnerer}, {Schuster}, \& {Usero}}]{leroy13a}
{Leroy} A.~K. {et~al.}, 2013, \aj, 146, 19

\bibitem[{{Longmore} {et~al}\mbox{.}(2013{\natexlab{a}}){Longmore}, {Bally},
  {Testi}, {Purcell}, {Walsh}, {Bressert}, {Pestalozzi}, {Molinari}, {Ott},
  {Cortese}, {Battersby}, {Murray}, {Lee}, {Kruijssen}, {Schisano}, \&
  {Elia}}]{longmore13a}
{Longmore} S.~N. {et~al.}, 2013{\natexlab{a}}, \mnras, 429, 987

\bibitem[{{Longmore} {et~al}\mbox{.}(2013{\natexlab{b}}){Longmore},
  {Kruijssen}, {Bally}, {Ott}, {Testi}, {Rathborne}, {Bastian}, {Bressert},
  {Molinari}, {Battersby}, \& {Walsh}}]{longmore13b}
{Longmore} S.~N. {et~al.}, 2013{\natexlab{b}}, \mnras, 433, L15

\bibitem[{{Longmore} {et~al}\mbox{.}(2012){Longmore}, {Rathborne}, {Bastian},
  {Alves}, {Ascenso}, {Bally}, {Testi}, {Longmore}, {Battersby}, {Bressert},
  {Purcell}, {Walsh}, {Jackson}, {Foster}, {Molinari}, {Meingast}, {Amorim},
  {Lima}, {Marques}, {Moitinho}, {Pinhao}, {Rebordao}, \&
  {Santos}}]{longmore12a}
{Longmore} S.~N. {et~al.}, 2012, \apj, 746, 117

\bibitem[{{Mac Low}(1999)}]{mac-low99b}
{Mac Low} M., 1999, \apj, 524, 169

\bibitem[{{Martini} {et~al}\mbox{.}(2003){Martini}, {Regan}, {Mulchaey}, \&
  {Pogge}}]{martini03a}
{Martini} P., {Regan} M.~W., {Mulchaey} J.~S., {Pogge} R.~W., 2003, \apj, 589,
  774

\bibitem[{{Meidt} {et~al}\mbox{.}(2013){Meidt}, {Schinnerer},
  {Garc{\'{\i}}a-Burillo}, {Hughes}, {Colombo}, {Pety}, {Dobbs}, {Schuster},
  {Kramer}, {Leroy}, {Dumas}, \& {Thompson}}]{meidt13a}
{Meidt} S.~E. {et~al.}, 2013, \apj, 779, 45

\bibitem[{{Meijerink} \& {Spaans}(2005)}]{meijerink05a}
{Meijerink} R., {Spaans} M., 2005, \aap, 436, 397

\bibitem[{{Meijerink} {et~al}\mbox{.}(2011){Meijerink}, {Spaans}, {Loenen}, \&
  {van der Werf}}]{meijerink11a}
{Meijerink} R., {Spaans} M., {Loenen} A.~F., {van der Werf} P.~P., 2011, \aap,
  525, A119+

\bibitem[{{Mills} {et~al}\mbox{.}(2015){Mills}, {Butterfield}, {Ludovici},
  {Lang}, {Ott}, {Morris}, \& {Schmitz}}]{mills15a}
{Mills} E.~A.~C., {Butterfield} N., {Ludovici} D.~A., {Lang} C.~C., {Ott} J.,
  {Morris} M.~R., {Schmitz} S., 2015, \apj~in~press, arXiv:1503.08137

\bibitem[{{Molina} {et~al}\mbox{.}(2012){Molina}, {Glover}, {Federrath}, \&
  {Klessen}}]{molina12a}
{Molina} F.~Z., {Glover} S.~C.~O., {Federrath} C., {Klessen} R.~S., 2012,
  \mnras, 423, 2680

\bibitem[{{Molinari} {et~al}\mbox{.}(2011){Molinari}, {Bally},
  {Noriega-Crespo}, {Compi{\`e}gne}, {Bernard}, {Paradis}, {Martin}, {Testi},
  {Barlow}, {Moore}, {Plume}, {Swinyard}, {Zavagno}, {Calzoletti}, {Di
  Giorgio}, {Elia}, {Faustini}, {Natoli}, {Pestalozzi}, {Pezzuto},
  {Piacentini}, {Polenta}, {Polychroni}, {Schisano}, {Traficante}, {Veneziani},
  {Battersby}, {Burton}, {Carey}, {Fukui}, {Li}, {Lord}, {Morgan}, {Motte},
  {Schuller}, {Stringfellow}, {Tan}, {Thompson}, {Ward-Thompson}, {White}, \&
  {Umana}}]{molinari11a}
{Molinari} S. {et~al.}, 2011, \apjl, 735, L33

\bibitem[{{Montenegro}, {Yuan} \& {Elmegreen}(1999){Montenegro}, {Yuan}, \&
  {Elmegreen}}]{montenegro99a}
{Montenegro} L.~E., {Yuan} C., {Elmegreen} B.~G., 1999, \apj, 520, 592

\bibitem[{{Nesvadba} {et~al}\mbox{.}(2011){Nesvadba}, {Boulanger}, {Lehnert},
  {Guillard}, \& {Salome}}]{nesvadba11a}
{Nesvadba} N.~P.~H., {Boulanger} F., {Lehnert} M.~D., {Guillard} P., {Salome}
  P., 2011, \aap, 536, L5

\bibitem[{{Offner}, {Hansen} \& {Krumholz}(2009){Offner}, {Hansen}, \&
  {Krumholz}}]{offner09b}
{Offner} S.~S.~R., {Hansen} C.~E., {Krumholz} M.~R., 2009, \apjl, 704, L124

\bibitem[{{Offner} {et~al}\mbox{.}(2008){Offner}, {Krumholz}, {Klein}, \&
  {McKee}}]{offner08a}
{Offner} S.~S.~R., {Krumholz} M.~R., {Klein} R.~I., {McKee} C.~F., 2008, \aj,
  136, 404

\bibitem[{{Ostriker}, {McKee} \& {Leroy}(2010){Ostriker}, {McKee}, \&
  {Leroy}}]{ostriker10a}
{Ostriker} E.~C., {McKee} C.~F., {Leroy} A.~K., 2010, \apj, 721, 975

\bibitem[{{Ostriker} \& {Shetty}(2011)}]{ostriker11a}
{Ostriker} E.~C., {Shetty} R., 2011, \apj, 731, 41

\bibitem[{{Ostriker}, {Stone} \& {Gammie}(2001){Ostriker}, {Stone}, \&
  {Gammie}}]{ostriker01a}
{Ostriker} E.~C., {Stone} J.~M., {Gammie} C.~F., 2001, \apj, 546, 980

\bibitem[{{Padoan} {et~al}\mbox{.}(2014){Padoan}, {Federrath}, {Chabrier},
  {Evans}, {Johnstone}, {J{\o}rgensen}, {McKee}, \& {Nordlund}}]{padoan14a}
{Padoan} P., {Federrath} C., {Chabrier} G., {Evans}, II N.~J., {Johnstone} D.,
  {J{\o}rgensen} J.~K., {McKee} C.~F., {Nordlund} {\AA}., 2014, Protostars and
  Planets VI, 77

\bibitem[{{Padoan}, {Haugb{\o}lle} \& {Nordlund}(2012){Padoan}, {Haugb{\o}lle},
  \& {Nordlund}}]{padoan12a}
{Padoan} P., {Haugb{\o}lle} T., {Nordlund} {\AA}., 2012, \apjl, 759, L27

\bibitem[{{Padoan} {et~al}\mbox{.}(2001){Padoan}, {Juvela}, {Goodman}, \&
  {Nordlund}}]{padoan01a}
{Padoan} P., {Juvela} M., {Goodman} A.~A., {Nordlund} {\AA}., 2001, \apj, 553,
  227

\bibitem[{{Padoan}, {Nordlund} \& {Jones}(1997){Padoan}, {Nordlund}, \&
  {Jones}}]{padoan97a}
{Padoan} P., {Nordlund} A., {Jones} B.~J.~T., 1997, \mnras, 288, 145

\bibitem[{{Papadopoulos}(2010)}]{papadopoulos10a}
{Papadopoulos} P.~P., 2010, \apj, 720, 226

\bibitem[{{Pettitt} {et~al}\mbox{.}(2014){Pettitt}, {Dobbs}, {Acreman}, \&
  {Price}}]{pettitt14a}
{Pettitt} A.~R., {Dobbs} C.~L., {Acreman} D.~M., {Price} D.~J., 2014, \mnras,
  444, 919

\bibitem[{{Piontek} \& {Ostriker}(2004)}]{piontek04a}
{Piontek} R.~A., {Ostriker} E.~C., 2004, \apj, 601, 905

\bibitem[{{Piontek} \& {Ostriker}(2007)}]{piontek07a}
{Piontek} R.~A., {Ostriker} E.~C., 2007, \apj, 663, 183

\bibitem[{{Purcell} {et~al}\mbox{.}(2012){Purcell}, {Longmore}, {Walsh},
  {Whiting}, {Breen}, {Britton}, {Brooks}, {Burton}, {Cunningham}, {Green},
  {Harvey-Smith}, {Hindson}, {Hoare}, {Indermuehle}, {Jones}, {Lo}, {Lowe},
  {Phillips}, {Thompson}, {Urquhart}, {Voronkov}, \& {White}}]{purcell12a}
{Purcell} C.~R. {et~al.}, 2012, \mnras, 426, 1972

\bibitem[{{Rathborne} {et~al}\mbox{.}(2015){Rathborne}, {Longmore}, {Jackson},
  {Alves}, {Bally}, {Bastian}, {Contreras}, {Foster}, {Garay}, {Kruijssen},
  {Testi}, \& {Walsh}}]{rathborne15a}
{Rathborne} J.~M. {et~al.}, 2015, \apj, 802, 125

\bibitem[{{Rathborne} {et~al}\mbox{.}(2014{\natexlab{a}}){Rathborne},
  {Longmore}, {Jackson}, {Foster}, {Contreras}, {Garay}, {Testi}, {Alves},
  {Bally}, {Bastian}, {Kruijssen}, \& {Bressert}}]{rathborne14b}
{Rathborne} J.~M. {et~al.}, 2014{\natexlab{a}}, \apj, 786, 140

\bibitem[{{Rathborne} {et~al}\mbox{.}(2014{\natexlab{b}}){Rathborne},
  {Longmore}, {Jackson}, {Kruijssen}, {Alves}, {Bally}, {Bastian}, {Contreras},
  {Foster}, {Garay}, {Testi}, \& {Walsh}}]{rathborne14a}
{Rathborne} J.~M. {et~al.}, 2014{\natexlab{b}}, \apjl, 795, L25

\bibitem[{{Renaud}, {Kraljic} \& {Bournaud}(2012){Renaud}, {Kraljic}, \&
  {Bournaud}}]{renaud12a}
{Renaud} F., {Kraljic} K., {Bournaud} F., 2012, \apjl, 760, L16

\bibitem[{{Rodriguez-Fernandez} \& {Combes}(2008)}]{rodriguez-fernandez08a}
{Rodriguez-Fernandez} N.~J., {Combes} F., 2008, \aap, 489, 115

\bibitem[{{Rosolowsky} \& {Blitz}(2005)}]{rosolowsky05a}
{Rosolowsky} E., {Blitz} L., 2005, \apj, 623, 826

\bibitem[{{Rosolowsky} {et~al}\mbox{.}(2008){Rosolowsky}, {Pineda}, {Foster},
  {Borkin}, {Kauffmann}, {Caselli}, {Myers}, \& {Goodman}}]{rosolowsky08a}
{Rosolowsky} E.~W., {Pineda} J.~E., {Foster} J.~B., {Borkin} M.~A., {Kauffmann}
  J., {Caselli} P., {Myers} P.~C., {Goodman} A.~A., 2008, \apjs, 175, 509

\bibitem[{{Saintonge} {et~al}\mbox{.}(2012){Saintonge}, {Tacconi}, {Fabello},
  {Wang}, {Catinella}, {Genzel}, {Graci{\'a}-Carpio}, {Kramer}, {Moran},
  {Heckman}, {Schiminovich}, {Schuster}, \& {Wuyts}}]{saintonge12a}
{Saintonge} A. {et~al.}, 2012, \apj, 758, 73

\bibitem[{{Sakamoto} {et~al}\mbox{.}(1999){Sakamoto}, {Okumura}, {Ishizuki}, \&
  {Scoville}}]{sakamoto99a}
{Sakamoto} K., {Okumura} S.~K., {Ishizuki} S., {Scoville} N.~Z., 1999, \apj,
  525, 691

\bibitem[{{Salim}, {Federrath} \& {Kewley}(2015){Salim}, {Federrath}, \&
  {Kewley}}]{salim15a}
{Salim} D.~M., {Federrath} C., {Kewley} L.~J., 2015, \apjl, 806, L36

\bibitem[{{Salpeter}(1955)}]{salpeter55a}
{Salpeter} E.~E., 1955, \apj, 121, 161

\bibitem[{{Sandstrom} {et~al}\mbox{.}(2010){Sandstrom}, {Krause}, {Linz},
  {Schinnerer}, {Dumas}, {Meidt}, {Rix}, {Sauvage}, {Walter}, {Kennicutt},
  {Calzetti}, {Appleton}, {Armus}, {Beir{\~a}o}, {Bolatto}, {Brandl},
  {Crocker}, {Croxall}, {Dale}, {Draine}, {Engelbracht}, {Gil de Paz},
  {Gordon}, {Groves}, {Hao}, {Helou}, {Hinz}, {Hunt}, {Johnson}, {Koda},
  {Leroy}, {Murphy}, {Rahman}, {Roussel}, {Skibba}, {Smith}, {Srinivasan},
  {Vigroux}, {Warren}, {Wilson}, {Wolfire}, \& {Zibetti}}]{sandstrom10a}
{Sandstrom} K. {et~al.}, 2010, \aap, 518, L59

\bibitem[{{Sani} {et~al}\mbox{.}(2012){Sani}, {Davies}, {Sternberg},
  {Graci{\'a}-Carpio}, {Hicks}, {Krips}, {Tacconi}, {Genzel}, {Vollmer},
  {Schinnerer}, {Garc{\'{\i}}a-Burillo}, {Usero}, \& {Orban de
  Xivry}}]{sani12a}
{Sani} E. {et~al.}, 2012, \mnras, 424, 1963

\bibitem[{{Schruba} {et~al}\mbox{.}(2011){Schruba}, {Leroy}, {Walter},
  {Bigiel}, {Brinks}, {de Blok}, {Dumas}, {Kramer}, {Rosolowsky}, {Sandstrom},
  {Schuster}, {Usero}, {Weiss}, \& {Wiesemeyer}}]{schruba11a}
{Schruba} A. {et~al.}, 2011, \aj, 142, 37

\bibitem[{{Shakura} \& {Sunyaev}(1973)}]{shakura73a}
{Shakura} N.~I., {Sunyaev} R.~A., 1973, \aap, 24, 337

\bibitem[{{Shetty} \& {Ostriker}(2012)}]{shetty12a}
{Shetty} R., {Ostriker} E.~C., 2012, \apj, 754, 2

\bibitem[{{Shu}(1992)}]{shu92a}
{Shu} F.~H., 1992, {Physics of Astrophysics, Vol. II}. University Science Books

\bibitem[{{\noopsort{Silva}da Silva}, {Fumagalli} \&
  {Krumholz}(2012){\noopsort{Silva}da Silva}, {Fumagalli}, \&
  {Krumholz}}]{da-silva12a}
{\noopsort{Silva}da Silva} R.~L., {Fumagalli} M., {Krumholz} M., 2012, \apj,
  745, 145

\bibitem[{{Sofue}(1995)}]{sofue95a}
{Sofue} Y., 1995, \pasj, 47, 527

\bibitem[{{Sofue} \& {Handa}(1984)}]{sofue84a}
{Sofue} Y., {Handa} T., 1984, \nat, 310, 568

\bibitem[{{Sormani}, {Binney} \& {Magorrian}(2015){Sormani}, {Binney}, \&
  {Magorrian}}]{sormani15a}
{Sormani} M.~C., {Binney} J., {Magorrian} J., 2015, \mnras, 449, 2421

\bibitem[{{Stone}, {Ostriker} \& {Gammie}(1998){Stone}, {Ostriker}, \&
  {Gammie}}]{stone98a}
{Stone} J.~M., {Ostriker} E.~C., {Gammie} C.~F., 1998, \apjl, 508, L99

\bibitem[{{Su}, {Slatyer} \& {Finkbeiner}(2010){Su}, {Slatyer}, \&
  {Finkbeiner}}]{su10a}
{Su} M., {Slatyer} T.~R., {Finkbeiner} D.~P., 2010, \apj, 724, 1044

\bibitem[{{Suwannajak}, {Tan} \& {Leroy}(2014){Suwannajak}, {Tan}, \&
  {Leroy}}]{suwannajak14a}
{Suwannajak} C., {Tan} J.~C., {Leroy} A.~K., 2014, \apj, 787, 68

\bibitem[{{Thompson} \& {Krumholz}(2014)}]{thompson14a}
{Thompson} T.~A., {Krumholz} M.~R., 2014, ArXiv e-prints

\bibitem[{{Thompson}, {Quataert} \& {Murray}(2005){Thompson}, {Quataert}, \&
  {Murray}}]{thompson05a}
{Thompson} T.~A., {Quataert} E., {Murray} N., 2005, \apj, 630, 167

\bibitem[{{Toomre}(1964)}]{toomre64a}
{Toomre} A., 1964, \apj, 139, 1217

\bibitem[{{Toomre}(1981)}]{toomre81a}
{Toomre} A., 1981, in Structure and Evolution of Normal Galaxies, {Fall} S.~M.,
  {Lynden-Bell} D., eds., pp. 111--136

\bibitem[{{Vazquez-Semadeni}(1994)}]{Vazquez-Semadeni94a}
{Vazquez-Semadeni} E., 1994, \apj, 423, 681

\bibitem[{{Walsh} {et~al}\mbox{.}(2011){Walsh}, {Breen}, {Britton}, {Brooks},
  {Burton}, {Cunningham}, {Green}, {Harvey-Smith}, {Hindson}, {Hoare},
  {Indermuehle}, {Jones}, {Lo}, {Longmore}, {Lowe}, {Phillips}, {Purcell},
  {Thompson}, {Urquhart}, {Voronkov}, {White}, \& {Whiting}}]{walsh11a}
{Walsh} A.~J. {et~al.}, 2011, \mnras, 416, 1764

\bibitem[{{Walsh}, {Myers} \& {Burton}(2004){Walsh}, {Myers}, \&
  {Burton}}]{walsh04a}
{Walsh} A.~J., {Myers} P.~C., {Burton} M.~G., 2004, \apj, 614, 194

\bibitem[{{Wang} {et~al}\mbox{.}(2012){Wang}, {Zhao}, {Mao}, \&
  {Rich}}]{wang12b}
{Wang} Y., {Zhao} H., {Mao} S., {Rich} R.~M., 2012, \mnras, 427, 1429

\bibitem[{{Willett} {et~al}\mbox{.}(2015){Willett}, {Schawinski}, {Simmons},
  {Masters}, {Skibba}, {Kaviraj}, {Melvin}, {Wong}, {Nichol}, {Cheung},
  {Lintott}, \& {Fortson}}]{willett15a}
{Willett} K.~W. {et~al.}, 2015, \mnras, in press, arXiv:150203444

\bibitem[{{Yusef-Zadeh} {et~al}\mbox{.}(2008){Yusef-Zadeh}, {Braatz}, {Wardle},
  \& {Roberts}}]{yusef-zadeh08a}
{Yusef-Zadeh} F., {Braatz} J., {Wardle} M., {Roberts} D., 2008, \apjl, 683,
  L147

\bibitem[{{Yusef-Zadeh} {et~al}\mbox{.}(2009){Yusef-Zadeh}, {Hewitt}, {Arendt},
  {Whitney}, {Rieke}, {Wardle}, {Hinz}, {Stolovy}, {Lang}, {Burton}, \&
  {Ramirez}}]{yusef-zadeh09a}
{Yusef-Zadeh} F. {et~al.}, 2009, \apj, 702, 178

\end{thebibliography}

\bsp

\label{lastpage}

\end{document}